\documentclass[sigconf, natbib=false]{acmart}

\usepackage{comment, graphicx, color, booktabs, algorithm, algpseudocode, xspace, listings, subcaption, multirow, enumitem, tikz, siunitx, amsmath, amsthm, xtab}
\usepackage{array}
\usepackage{threeparttable}
\usepackage{soul}
\usepackage{bbding} 
\usepackage{pifont}
\usepackage{pbox}
\usepackage[many]{tcolorbox}
\usepackage{balance}
\newtcolorbox[auto counter]{mybox}[2][]{enhanced jigsaw,  breakable, #1}

\graphicspath{{figure/}}

\usepackage[backend=biber, style=acmnumeric, natbib=true]{biblatex}
\renewbibmacro*{year}{%
  \iffieldundef{year}%
    {}
    {\printfield{year}}%
}

\newcommand{\tool}{\textsc{Hopper}\xspace}
\newcommand*\circled[1]{\tikz[baseline=(char.base)]{
            \node[shape=circle,fill,inner sep=0.8pt] (char) {\textcolor{white}{#1}};}}
\newcommand{\fn}[1]{\mbox{\textsc{#1}}}
\newcommand{\kw}[1]{\textbf{#1}\xspace}
\newcommand{\file}[1]{\textit{#1}\xspace}

\newcommand{\func}[1]{\mbox{\small{\texttt{#1}}}}
\newcommand{\tablength}{2pt}

\lstdefinelanguage{Hopper}{
  sensitive = true,
  keywords={call, load, update, assert, file},
  morekeywords=[2]={mut*, const*},
  comment=[l]{//},
  morestring=[b]',
  morestring=[b]"
}

\title{\tool: Interpretative Fuzzing for Libraries}

\author{Peng Chen}
\affiliation{
  \institution{Tencent Security Big Data Lab}
  \country{} 
}
\email{spinpx@gmail.com}

\author{Yuxuan Xie}
\affiliation{
  \institution{Tencent Security Big Data Lab}
  \country{} 
}
\email{ryanyxie@tencent.com}

\author{Yunlong Lyu}
\affiliation{
  \institution{Tencent Security Big Data Lab}
  \country{} 
}
\email{loydlv@tencent.com}

\author{Yuxiao Wang}
\affiliation{
  \institution{Tencent Security Big Data Lab}
  \country{} 
}
\email{yuxiaowang@tencent.com}

\author{Hao Chen}
\affiliation{
  \institution{University of California, Davis}
  \country{} 
}
\email{chen@ucdavis.edu}

\copyrightyear{2023}
\acmYear{2023}
\setcopyright{acmlicensed}
\acmConference[CCS '23]{Proceedings of the 2023 ACM SIGSAC Conference on Computer and Communications Security}{November 26--30, 2023}{Copenhagen, Denmark}
\acmBooktitle{Proceedings of the 2023 ACM SIGSAC Conference on Computer and Communications Security (CCS '23), November 26--30, 2023, Copenhagen, Denmark}
\acmPrice{15.00}
\acmDOI{10.1145/3576915.3616610}
\acmISBN{979-8-4007-0050-7/23/11}

\begin{document}

\begin{abstract}

Despite the fact that the state-of-the-art fuzzers can generate inputs efficiently, existing fuzz drivers still cannot adequately cover entries in libraries. Most of these fuzz drivers are crafted manually by developers, and their quality depends on the developers' understanding of the code. Existing works have attempted to automate the generation of fuzz drivers by learning API usage from code and execution traces. However, the generated fuzz drivers are limited to a few specific call sequences by the code being learned. To address these challenges, we present \tool, which can fuzz libraries without requiring any domain knowledge to craft fuzz drivers. It transforms the problem of library fuzzing into the problem of interpreter fuzzing. The interpreters linked against libraries under test can interpret the inputs that describe arbitrary API usage. To generate semantically correct inputs for the interpreter, \tool learns the intra- and inter-API constraints in the libraries and mutates the program with grammar awareness. We implemented \tool and evaluated its effectiveness on 11 real-world libraries against manually crafted fuzzers and other automatic solutions. Our results show that \tool greatly outperformed the other fuzzers in both code coverage and bug finding, having uncovered 25 previously unknown bugs that other fuzzers couldn't. Moreover, we have demonstrated that the proposed intra- and inter-API constraint learning methods can correctly learn constraints implied by the library and, therefore, significantly improve the fuzzing efficiency. The experiment results indicate that \tool is able to explore a vast range of API usages for library fuzzing out of the box.

\end{abstract}



\maketitle

\section{Introduction}
\label{sec:intro}

Fuzzing is one of the most popular techniques for finding software vulnerabilities. Fuzzers give the software a large number of random inputs and observe whether unexpected behaviors happen. Though the idea is simple, fuzzing has successfully been applied to testing various applications and found many bugs. In recent years, the techniques have been greatly improved due to the advent of grey-box fuzzing. Coverage-based grey-box fuzzers such as AFL~\cite{afl} and LibFuzzer~\cite{libfuzzer} mutate inputs to explore deeper program states without requiring knowledge about input format or program specifications. Constraint-based grey-box fuzzers~\cite{angora, matryoshka, greyone, qsym, driller, chen2022jigsaw}, as the state-of-the-art fuzzers, employ constraint-solving techniques to reach code branches that are guarded by complex constraints.

While grey-box fuzzing techniques greatly facilitate the generalization and automation of program fuzzing (i.e., fuzzing techniques that take ready-to-use binary programs as targets), replicating such success in library fuzzing (i.e., fuzzing techniques that take APIs as targets) is challenging. To use the state-of-the-art fuzzers for libraries, users have to manually craft a fuzz driver that consumes the type-agnostic inputs fed by fuzzers and transforms the byte stream into API arguments. However, writing high-quality fuzz drivers is difficult, as it is time-consuming and requires a deep understanding of the library. Consequently, most of the existing fuzz drivers cover only a small part of library APIs. The APIs, especially the rarely used ones, usually lack adequate testing. Library fuzzing is still struggling to be usable out-of-box and scalable due to the lack of an automated solution.

In library fuzzing, fuzz drivers need to use correct argument types when invoking APIs and satisfy both intra- and inter-API constraints. Otherwise, the fuzz drivers would crash unexpectedly. For instance, to test \lstinline{ares_send(ares_channel channel, char *qbuf, int qlen, ares_callback callback, void *arg)} from \file{c-ares}, a fuzz driver should initialize the first argument \verb|channel| with a call of \lstinline|ares_init(ares_channel *channelptr)| to meet the inter-API constraint, and then set the third argument \verb|qlen| to be the length of the second argument \verb|qbuf| and set the forth argument \verb|callback| to be a non-null function pointer with the type of \verb|ares_callback| to comply with the intra-API constraints. Any violation of these three constraints may result in spurious crashes. However, in practice, information about these constraints is either missing or scattered across library documents or comments, making it hard to collect them in a fully automatic way.

Recently, researchers have proposed learning-based~\cite{babic2019fudge, ispoglou2020fuzzgen, zhang2021apicraft, jung2021winnie} and model-based~\cite{green2022graphfuzz, lpm, fdp} methods for automatically generating fuzz drivers. Learning-based methods, such as FuzzGen\cite{ispoglou2020fuzzgen}, attempt to learn the correct usage of APIs from existing consumer code. However, this method fails when consumer code is unavailable, such as for new or work-in-progress libraries. Model-based methods, such as GraphFuzz~\cite{green2022graphfuzz}, ask users to provide specifications of the APIs under test, which requires domain-specific knowledge and significant human involvement. Furthermore, the quality of the fuzz drivers generated with these methods is largely affected by the external inputs (i.e., consumer code or user-provided expertise), which can be inaccurate or incomplete. For example, in our experiments, we found that some of the fuzz drivers provided by the authors of FuzzGen~\cite{ispoglou2020fuzzgen} and GraphFuzz~\cite{green2022graphfuzz} result in spurious crashes due to misuses in the consumer code and incorrect user-defined schemas (\autoref{sec:wrong_drivers} in Appendix). The fuzz drivers~\cite{fuzzgen_libvpx} generated by FuzzGen only cover 5 of 26 API functions in the \file{libvpx} decoding library and only support the \file{vp9} codec. Similarly, the specifications~\cite{sqlite3_gfuzz_schema} of \file{sqlite3} written by the authors of GraphFuzz do not even include commonly used API functions, such as \lstinline|sqlite3_exec| and \lstinline|sqlite3_complete|.

To address the aforementioned challenges, we present \tool to fuzz APIs without requiring external knowledge. Inspired by coverage-based fuzzing, which learns valid format from mutating random seeds~\cite{pulling_jepgs}, \tool learns potential usage of the APIs from mutating the composition of API calls and their arguments. If executing the mutated program triggers a new path or a new crash, \tool infers the intra- and inter-API constraints based on the dynamic feedback. To achieve this, we introduce a Domain-Specific Language (DSL) that describes arbitrary API usages. The DSL inputs can be interpreted with a lightweight interpreter linked against the library under test. In this way, we transform library fuzzing into interpreter fuzzing. The fuzzer is now responsible for generating programs encoded in the format of DSL to feed into the interpreter. Then the interpreter executes the programs to see if unexpected behavior happens. Thanks to the grammar-aware mutation and the inferred constraints, \tool is able to generate valid inputs that explore different API usage while excluding false positive crashes.

We implemented \tool and evaluated its effectiveness on 11 real-world libraries against manually crafted fuzzers (MCF) and other automatic solutions (i.e., FuzzGen and GraphFuzz). \autoref{tbl:cov} shows that \tool greatly outperformed the other fuzzers in both code coverage and bug finding. Notably, \tool improves the code coverage over MCFs in \file{cJSON} by 47.12\% in line coverage and 37.50\% in branch coverage and achieves higher coverage than the total of 9 fuzzers in \file{zlib}. In total, \tool found 25 new bugs in the libraries, and 17 of them have been confirmed, as shown in \autoref{tbl:bugs}.
Moreover, we have demonstrated that the proposed intra- and inter-API constraint learning methods can accurately learn constraints implied by the library, thereby significantly improving the fuzzing efficiency.

\section{Background}

\subsection{Library Fuzzing}

The impact of security vulnerabilities in libraries is significant due to their widespread use in various programs. However, testing libraries through program fuzzing alone is often inadequate.
Library APIs may be invoked by programs under complex path constraints or with specific arguments that make it difficult to exercise the APIs fully.
To address this issue, library fuzzing tools have been developed to specifically test libraries, with LibFuzzer being a common choice for such testing. Here are the required steps for effectively using LibFuzzer to fuzz libraries:

\begin{itemize} 
\item \textbf{Craft a fuzz drivers}. 
A fuzz driver is a program that describes the usage of library APIs, including a sequence of API calls and their arguments. A high-quality fuzz driver should provide an entry to explore as much code in the library as possible for thorough testing. 
Specific execution paths are not only determined by the arguments in API calls but also by invoking their related APIs. These related APIs may return values as arguments or affect the context for other APIs that rely on them (e.g., global values).
Hence, a deep understanding of the tested library is necessary for thorough testing. However, enumerating all valid API usages would be time-consuming and challenging. Therefore, fuzz drivers usually contain only a few common API usages.
As shown in \autoref{tbl:api_cov},  the MCFs of the listed 11 popular libraries only covered 18.58\% of the APIs, whereas the remaining 81.42\% uncovered APIs will escape being fuzzed. Even though some approaches have been proposed to automatically synthesize fuzz drivers, the coverage of APIs is still limited (e.g., GraphFuzz~\cite{green2022graphfuzz} achieved API coverage of 41.42\%).
Moreover, since different API usages have different search spaces for fuzzing, putting all of them in sequence would be inefficient. Instead, it would be more effective to write them into multiple fuzz drivers or execute them conditionally within a single fuzz driver. For example, in \autoref{fig:driver_example}, the fuzz driver calls the parsing function and then calls different printing functions according to the first 4 bytes of \verb|data|. 
    
\item \textbf{Specify the format of input}.
The blind byte stream generated by LibFuzzer makes it difficult to create structured inputs that satisfy intra-API constraints. To address this issue, we need to specify the input format and guide the fuzzer to generate arguments beyond byte arrays.
FuzzedDataProvider~\cite{fdp} is capable of dividing the fuzz input into multiple parts of various types, while libprotobuf-mutator\cite{lpm} can generate structured inputs based on provided grammars. 
However, the presence of intra-API constraints makes defining the potential argument range for fuzzing an enormous task. Hence, developers might opt to encode arguments as literal constants directly into fuzz drivers, such as the second argument of \func{cJSON\_ParseWithOpts} in \autoref{fig:driver_example}. Unfortunately, this approach can result in inadequate testing for API functions.
    
\end{itemize}

\begin{table}[t]
\setlength{\tabcolsep}{\tablength}
\caption{Number of unique APIs used in fuzz drivers. The second column is the total number of exported APIs in libraries.}
\resizebox{0.97\columnwidth}{!}{
\begin{threeparttable}
\begin{tabular}{lccccc}
\toprule
\textbf{Library} & \textbf{Total} & \textbf{MCFs} & \textbf{FuzzGen}$^*$ & \textbf{GraphFuzz} & \textbf{\tool} \\
\midrule
\file{cJSON}            & 78   & 6(7.69\%)    & 4(4.13\%)    & 40(51.28\%)   & 78(100\%)     \\
\file{c-ares}           & 60   & 13(21.67\%)  & -            & 20(33.33\%)   & 58(96.67\%)   \\
\file{libpng}           & 241  & 25(10.37\%)  & -            & 66(27.39\%)   & 233(96.69\%)  \\
\file{lcms}             & 283  & 10(3.53\%)   & -            & 38(13.43\%)   & 274(96.82\%)  \\
\file{libmagic}         & 18   & 4(22.22\%)   & -            & 10(55.56\%)   & 18(100\%)     \\
\file{libpcap}          & 89   & 8(8.99\%)    & -            & 29(32.58\%)   & 73(82.02\%)   \\
\file{zlib}             & 84   & 29(34.52\%)  & -            & 41(48.81\%)   & 78(92.86\%)   \\
\file{re2}              & 70   & 35(50.00\%)  & -            & 47(67.14\%)   & 69(98.57\%)   \\
\file{sqlite3}          & 279  & 15(5.38\%)   & -            & 74(26.52\%)   & 225(80.65\%)  \\
\file{libvpx}           & 26   & 7(26.92\%)   & 5(19.23\%)   & 17(65.38\%)   & 24(92.31\%)   \\
\file{libaom}           & 38   & 5(13.16\%)   & 7(18.42\%)   & 13(34.21\%)   & 35(92.11\%)   \\
\midrule
\textbf{Average}        & -    & \textbf{18.58\%}  & \textbf{13.93\%}  & \textbf{41.42\%}  & \textbf{93.52\%}    \\
\bottomrule
\end{tabular}
\footnotesize{$^*$The authors of FuzzGen released the fuzz drivers for \file{libvpx} and \file{libaom}, and the released code of FuzzGen is unable to run on the rest libraries, except for \file{cJSON}.}
\end{threeparttable}
}
\label{tbl:api_cov}
\end{table}

\subsection{Fuzzing Interpreters}
Grammar-aware grey-box fuzzing has succeeded in programs that parse the inputs, especially in interpreters~\cite{srivastava2021gramatron, wang2019superion, aschermann2019nautilus, blazytko2019grimoire, gross2018fuzzil, veggalam2016ifuzzer, lee2020montage, chen2021one}.  The success of fuzzing interpreters is attributed to the following two key techniques.
\begin{itemize}
    \item \textbf{Grammar-aware input mutation}. Blindly mutated inputs are likely to be rejected by the parsing procedure, while structured inputs can reach deeper paths. Grammar-aware mutations parse the inputs as intermediate representations (IRs) based on their encoding grammar and mutate the IRs while adhering to constraints.
    For example, Superion~\cite{wang2019superion} uses an abstract syntax tree as the IR and conducts three main operators on the tree for mutation: replace a node in the tree with a newly created sub-tree, splice two different trees, and minimize trees without affecting execution. 
   
    \item \textbf{Coverage guided fuzzing}.  Guided by the coverage feedback, the fuzzer keeps inputs that trigger new paths and mutates them further to enter deeper branches or find new bugs. By incorporating grammar-aware mutation strategies, the fuzzer can effectively synthesize valid test inputs that cover more branches and trigger new bugs.
    
\end{itemize}

We observe that constructing fuzz drivers for libraries resembles implementing an interpreter for the inputs, and fuzzing the interpreter is equivalent to fuzzing the library under the hood. However, MCFs interpret type-agnostic byte inputs into typed values to feed limited sequences of API calls, which only partially explore the libraries. If we extend the interpreter to accept any API usages as input, a grammar-aware coverage-based fuzzer can efficiently generate inputs for the interpreter and thoroughly explore the libraries.

\section{Design}

\subsection{Overview}

\begin{figure}[t]
  \centering \includegraphics[width=1\linewidth]{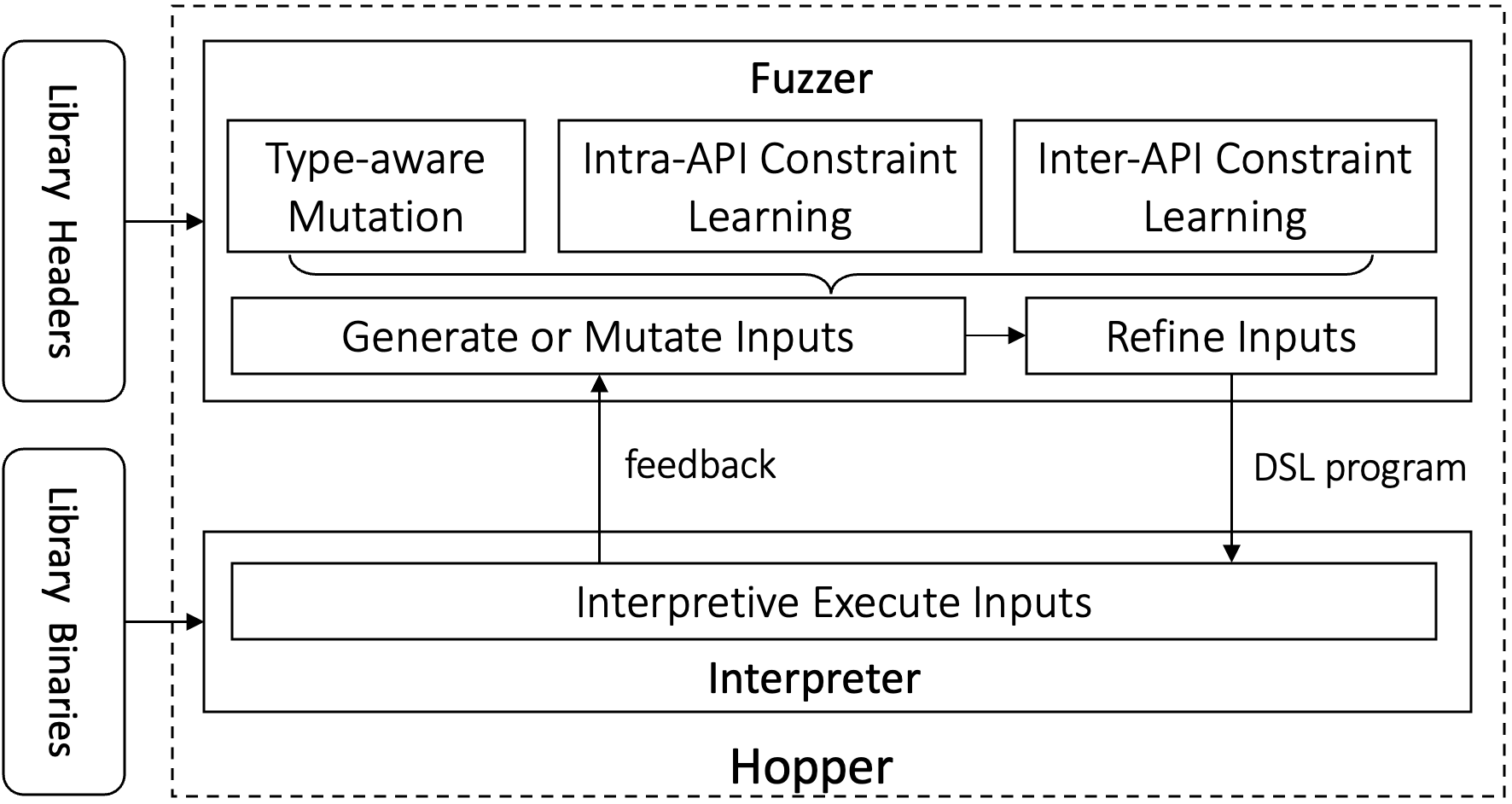}
  \caption{Overview of \tool 's architecture.}
  \label{fig:overview}
\end{figure}

\tool transforms the problem of library fuzzing into the problem of interpreter fuzzing.
At a high level, \tool consists of two main components: a grammar-aware fuzzer that generates inputs encoded in a DSL format, and a lightweight interpreter that executes these inputs, as shown in \autoref{fig:overview}.

The fuzzer produces high-quality inputs for invoking library APIs. It first extracts function signatures and type definitions from the library's header files, which provide valuable information on potential relations between API functions and argument types. By leveraging this information, \tool generates a sequence of API calls through random combinations of API functions and arguments. 
At the early stage, the generated sequences may be invalid and terminate with shallow paths in the APIs' code. However, thanks to coverage guidance, \tool keeps the inputs that explore new branches as seeds and mutate them further. Ultimately, the generated sequences become valid and their execution can reach deeper code.
Completely random fuzzing wastes enormous time in generating illegal and inefficient inputs. To accelerate the process of evolution, \tool performs type-aware argument mutation (\autoref{sec:type_mutate}) and learns intra- and inter-API constraints (\autoref{sec:constraint}). Additionally, \tool minimizes input size (\autoref{sec:minimize}) to reduce overhead.

The interpreter consumes the inputs and invokes the library APIs. The libraries under test are linked against the interpreter during the compiling stage. Before linking, \tool instruments the binaries to capture internal states during execution, e.g., code coverage. When new input arrives, the interpreter parses the input according to the syntax of DSL and interprets the statements orderly based on their semantics. 

\subsection{DSL and Input Interpretation}
\label{sec:runtime}
Fuzz drivers are usually developed with the same programming language as the library's specification. Therefore, for compiled language like C/C++, fuzz drivers are crafted and compiled before fuzzing begins. Only the input bytes are changed during each round of fuzzing. To replace the sequence of API calls during fuzzing while avoiding compilation overhead, we introduce a Domain Specific Language (DSL) and a lightweight interpreter in \tool for accelerating the entire process.

\subsubsection{DSL}
\label{sec:dsl}
For the sake of generality, the interpreter in \tool takes \textit{DSL programs} as input. The grammar of \tool DSL is listed in \autoref{fig:dsl} in the appendix. 
Each \textit{DSL program} comprises \textit{statements} as its most fundamental components, and each statement has a unique index that its successors can reference. 
We categorize common fuzzing behaviors found in MCFs into five \textit{statement} types in our DSL. 

\begin{itemize}
    \item A \textbf{load statement} defines the type information and literal representation of a value. Strong typing enables the generation of input data directly in a specific type, which eliminates the requirement of converting it from a type-agnostic byte stream.
    \item A \textbf{call statement} invokes a specific API function in libraries by providing the function name and a list of arguments, where each argument refers to the value defined in a previous load statement.
    \item An \textbf{update statement} overwrites the value returned by a call statement at runtime.
    \item An \textbf{assert statement} checks a call's return value at runtime. If the assertion fails (e.g., the caller needs to dereference the return value but the value is null), the program exits immediately.
    \item A \textbf{file statement} specifies a valid file resource for I/O operations. If the statement is used for reading, a sequence of random bytes is filled in the file for runtime reading.
\end{itemize}

To keep the grammar simple, our DSL does not support conditional statements. Some fuzz drivers use conditional statements to choose different API functions or arguments. By contrast, \tool generates different DSL programs to enumerate those API functions or arguments. For example, \autoref{fig:driver_example} is a real-world fuzz driver written by the developers of \file{cJSON}, and \autoref{fig:dsl_example} is a program written in DSL that covers a path in \autoref{fig:driver_example}.

\subsubsection{Interpreter}
\label{sec:interpreter}
The interpreter parses DSL programs, executes the statements gracefully, and monitors the states of the program after executing each statement.
In order to invoke the library APIs, \tool links the interpreter against the libraries under test at compiling stage. It also constructs a table that correlates each function's name with its corresponding caller based on the header files of the libraries. During program execution, the caller casts values to the types of arguments it needs and then invokes the referenced function with the arguments.
Before linking to the interpreter, \tool instruments the library binaries with code that counts branches, and hooks compare instructions and resource management functions (e.g., \verb|malloc|, \verb|free| and \verb|fopen|). The interpreter then collects the following feedback at runtime.

\begin{itemize}
    \item \textbf{Optional Branch Tracking.}  It is unnecessary to track branches of all the library API calls in the DSL program every time, so \tool defines a global flag to guard the branch tracking code. The value of the flag is determined by the DSL input. When calling an API function that ends with a question mark (e.g., Line 13 in \autoref{fig:dsl_example}), the interpreter enables branch tracking for that call. 
    \item \textbf{Context-sensitive Code Coverage.} To distinguish the same branches visited by different APIs, the interpreter sets the hash of the current API function's name as the context. The instrumentation code reads the context and calculates an exclusive code trace for each API.
    \item \textbf{Overflow Detection.} If the statement loads a variable-sized value (e.g., an array), the interpreter of \tool stores them in a memory arena and appends a canary right after it to detect possible buffer overflow.
    \item \textbf{Use-after-free Detection.} The interpreter maintains a set of memory chunks allocated by \func{malloc} and released by \func{free} through instrumentation. For each pointer used in function arguments, if the memory chunk pointed by this pointer is freed, the interpreter exits the program immediately to avoid the use-after-free issues.
    \item \textbf{Comparison Hooking.} The interpreter collects the parameters used in comparison instructions and functions to guide the fuzzer to solve magic bytes.
\end{itemize}

The fuzz drivers utilized by LibFuzzer should have no side effects since the code runs in a loop in the same process. However, it is difficult to generate a program that can reset all resources before exiting. \tool's interpreter solves this problem by running each input in an individual process. Once a DSL program terminates, the operating system destroys the process and releases all allocated resources. This allows the interpreter to execute DSL programs continuously without the need to release resources.

\begin{figure}[t]
\begin{lstlisting}[language=C]
int LLVMFuzzerTestOneInput(const uint8_t* data, size_t size) {
  cJSON *json;
  size_t offset = 4;
  unsigned char *copied;
  char *printed_json = NULL;
  int minify = data[0] == '1' ? 1 : 0;
  int require_termination = data[1] == '1' ? 1 : 0;
  int formatted = data[2 ]== '1' ? 1 : 0;
  int buffered = data[3] == '1' ? 1 : 0;
  if (size <= offset) return 0;
  json = cJSON_ParseWithOpts((const char*)data + offset, NULL, require_termination);
  if (json == NULL) return 0;
  if(buffered) {
    printed_json=cJSON_PrintBuffered(json, 1, formatted);
  } else {
    if(formatted)  printed_json = cJSON_Print(json);
    else printed_json = cJSON_PrintUnformatted(json);
  }
  if(printed_json != NULL)  free(printed_json);
  if(minify) { ... }
  cJSON_Delete(json);
  return 0;
}
\end{lstlisting}
\caption{Fuzz driver written by the developer of cJSON. The code is in cjson\_read\_fuzzer.c, which is used by OSS-Fuzz~\cite{serebryany2017oss}. }
\label{fig:driver_example}
\end{figure}

\begin{figure}[t]
\begin{lstlisting}[language=Hopper, numbers=none]
<0>  load Vec<char>= vec(32)["GXsAAAAAAAAAo9tsrXXoqw57jwAAAAAAAAARNk+1AAA="]
<1>  load char* = &<0>
<2>  load char** = null
<3>  load int = 0
<4>  call cJSON_ParseWithOpts (<1>, <2>, <3>)
<5>  assert non_null(<4>)
<6>  load cJSON = { next: null, prev:  null, child: null, type_: 8, valuestring: null, valueint: 12345, valuedouble: 0.2771, string: null }
<7>  update <4>[0.child] = <6>
<8>  load Vec<char> = vec(7)[54, 52, -68, -43, 1, 122, 0]
<9>  load char* = &<8>
<10> call cJSON_AddFalseToObject (<4>, <9>)
<11> load int = 1
<12> load int = 0
<13> call cJSON_PrintBuffered ? (<4>, <11>, <12>)
\end{lstlisting}
\caption{Example program in the format of \tool DSL.  }
\label{fig:dsl_example}
\end{figure}

\subsection{Grammar-aware Input Fuzzing}

\begin{figure*}[t]
  \centering \includegraphics[width=1\linewidth]{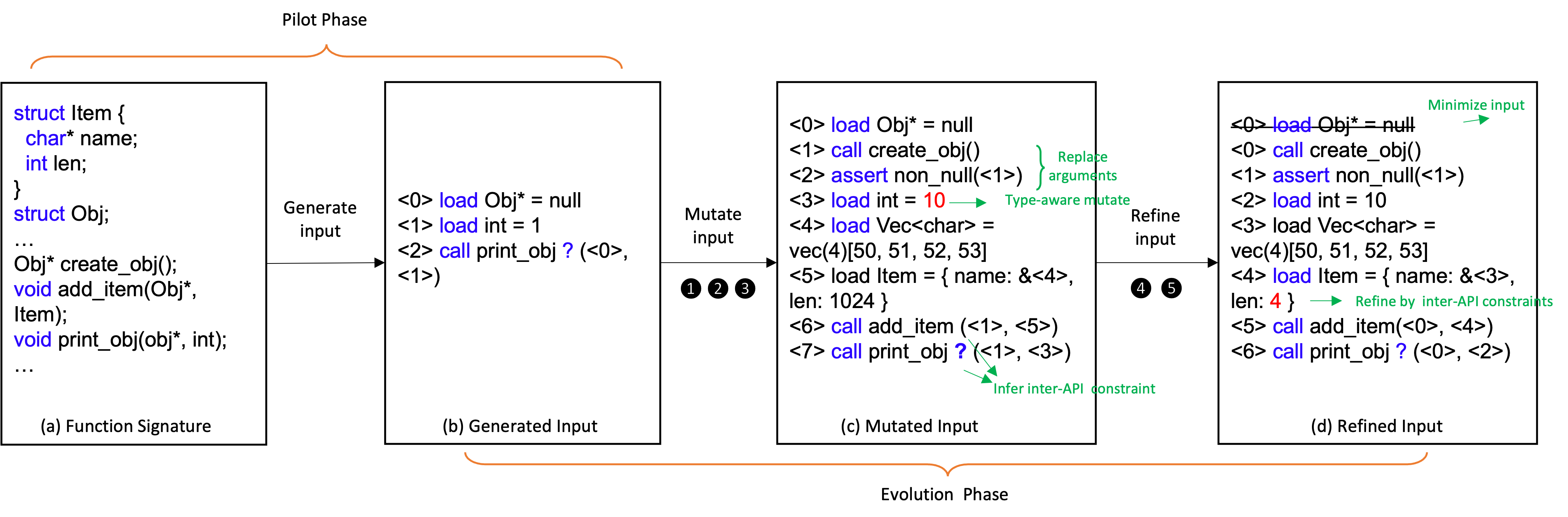}
  \caption{The workflow of input generation in \tool.}
  \label{fig:workflow}
\end{figure*}

To find bugs in libraries rather than interpreters, \tool takes a different approach compared to other grammar-aware fuzzers. While other fuzzers traverse all possible combinations of syntaxes based on input grammar, \tool instead focuses on generating various effective API calls.
This process involves two phases, as shown in \autoref{fig:workflow}. First, \tool generates inputs based on the information available in function signatures to initialize a seed pool. Second, \tool selects inputs in the seed pool and mutates them with the guidance of coverage feedback.

\subsubsection*{Pilot Phase}
\label{sec:pilot}
In the pilot phase, \tool draws skeletons of the inputs and infers constraints. Initially, the seed pool contains no inputs. Therefore, \tool tries to generate simple inputs for each API function based on their signatures and learn constraints from them (detailed in \autoref{sec:constraint}). 
To accomplish this, \tool selects an API function in the library as a target and attempts to randomly generate a call statement for it. This includes generating arguments and inserting related calls that introduce the necessary context. These statements finally form an input, which is then executed by the interpreter. If the input triggers a new path in the libraries without crashing, \tool saves it into the seed pool for further mutation. To prevent irrelevant coverage feedback from other calls, \tool tracks the coverage of the target call only.
 
\autoref{alg:callgen} describes the procedure of generating a call statement. Each argument is generated according to its type in the function signature using one of the following three operators:
\begin{itemize}
\item \tool chooses an existing statement whose type matches the argument\footnote{\tool treats const and non-const types equally when comparing types. }. 
\item The argument is obtained by inserting a new API invocation. \tool randomly selects an API function whose return type matches the argument, and generates a call statement for it recursively. The new call statements are placed ahead of the current one. Additionally, an assertion statement indicating whether the call statement runs successfully is added after the new call statement (e.g., a non-null assertion statement is inserted after a pointer-type returned call statement).
\item  A load statement with a typed value created from scratch is used. 
\end{itemize}

\tool also attempts to affect the execution of the target call by inserting other API calls to change the internal states of the program. It prioritizes API functions with non-primitive argument types that overlap with those of the target call and reuses the overlapping arguments as much as possible.
To prevent the program from becoming overly complex, \tool stops generating new calls if the length of statements exceeds a specific threshold or the recursion depth becomes too high.
Take the program in \autoref{fig:dsl_example} as an example. To generate the program with \verb|cJSON_PrintBuffered| (Line 13), \tool randomly chooses the return value of \verb|cJSON_ParseWithOpts| (Line 4) and generate two integer value (Line 11 and 12) as its arguments. In addition, it creates a call for \verb|cJSON_AddFalseToObject| that may modify existing arguments (Line 10).
 
\begin{algorithm}
    \begin{algorithmic}[1]
    \Function{GenerateCall}{$program, sig$}
    \State $args \gets$ empty list
    \ForAll{$arg\_type \in sig.arg\_types$}
        \If {\fn{Rand}() $< THRESHOLD$}  \Comment{Use existing statement as argument}
            \State $index \gets$ Randomly choose a statement with $arg\_type$ in the $program$.
            \If {$index$ exists}
                \State $args \gets args  \cup index$
                \State \kw{continue}
            \EndIf
        \EndIf
        \If {\fn{Rand}() $< THRESHOLD$ \kw{and} $arg\_type$ is not primitive} \Comment{Generate a new call statement as argument}
            \State $f \gets $ Randomly choose an API function that returns $arg\_type$ .
            \If {$f$ exists}
                \State $call \gets$ \fn{GenerateCall}($program, f$)
                \State $index \gets$ \fn{InsertStmt}($program, call$)
                \State $check \gets$ \fn{GenerateAssert}($program, index$)
                \State \fn{InsertStmt}($program, check$)
                \State $args \gets args \cup index$
                \State \kw{continue}
            \EndIf
        \EndIf
        \State $load \gets$ \fn{GenerateLoad}($program, arg\_type$) \Comment{Generate a new load statement as argument}
        \State $index \gets$ \fn{InsertStmt}($program, load$)
        \State $args \gets args  \cup index$
    \EndFor
    \If {\fn{Rand}() $< THRESHOLD$} \Comment{Generate a new call statement that may affect the execution of $call$}
        \State $f \gets $ Randomly choose an API function. \Comment{Prefer API functions whose argument types overlap with $args$}
        \State $call \gets$ \fn{GenerateCall}($program, f$)
        \State \fn{InsertStmt}($program, call$)
    \EndIf
    \State $call \gets$ \fn{CreateCallStmt}($sig, args$)
    \State \kw{return} call
    \EndFunction
\end{algorithmic}
\caption{Randomly generate a call statement based on its function signature. If the length of statements in the $program$ exceeds a certain length, stop generating new calls recursively.}
\label{alg:callgen}
\end{algorithm}

\subsubsection*{Evolution Phase}
\label{sec:evolution}
After running a certain number of rounds, \tool enters the evolution phase, which aims to build more complex programs based on the skeleton inputs. To achieve this, \tool selects an input from the seed pool and randomly mutates the statements based on their types. Guided by the execution branch coverage, \tool keeps the mutated inputs that explore deeper code of libraries, which are more likely to find bugs. Overall, \tool mutates program statements in five steps. 
\circled{1} \tool selects an input from the seed pool with priority. Fresh seeds are given higher priority to be chosen, as they are more likely to reach deeper paths. 
\circled{2} \tool chooses statements to mutate from the input according to their weights. The assert, file, and update statements are weighted as zero and not mutated, while the weights of load and call statements are determined by their complexity. 
\circled{3} \tool mutates the load statements and call statements by their corresponding strategies, as described in \autoref{sec:call_mutate} and \autoref{sec:type_mutate} respectively.
\circled{4} \tool refines the input to use APIs correctly based on the constraints learned during fuzzing, as described in \autoref{sec:constraint}.
\circled{5} \tool minimizes the input to remove redundant statements that have no effect on reaching the path yet increase the search space of mutation, as described in \autoref{sec:minimize}.

\subsubsection{Call Statement Mutation}
\label{sec:call_mutate}
\tool adopts the following mutation strategies for call statements:    
\begin{itemize}
    \item Replace one of the arguments with a new argument that retains the same type (Line 4 to Line 24 in \autoref{alg:callgen}). 
    \item Insert a new call before the target call (Line 27 to Line 29 in \autoref{alg:callgen}). The inserted call may modify the values of the target call's arguments or change the global states in libraries. \tool determines the effectiveness of the inserted calls through branch feedback, which is detailed in \autoref{sec:interapi}.
    \item Update the return value of the call. An update statement is inserted after the call to partially overwrite the return value with a new value.
\end{itemize}

\subsubsection{Type-aware Value Mutation}
\label{sec:type_mutate}

As libraries often use various argument types, including custom composite types, \tool needs to generate appropriate types for such arguments when invoking the corresponding APIs. The same is true for value mutation, where applying mutations according to the value's type can more effectively explore new states.
To achieve this, \tool parses type definitions (e.g., \verb|struct|, \verb|enum| and \verb|union|) and type aliases in header files recursively. It then generates new typed values using the following rules:

\begin{itemize}

\item \textbf{Primitive Types.} Almost all primitive types are numerical types. Thus, \tool generates numbers within a small range of uniform distribution. Additionally, \tool applies one of four mutations to the primitive values.
(1) Set an interesting value (e.g., 0x80000000 for int); 
(2) Flip a bit or a byte; 
(3) Add or subtract a small number; 
(4) Set to a literal collected from a comparison instruction, if the value is used as one of the operands.
 
\item \textbf{Array.} Based on the array's length and element type, \tool generates a sequence of elements. If the length is variable, \tool first randomly chooses a length. During mutation, \tool selects one or more elements in the array and mutates them respectively. In addition, \tool can resize the array by inserting or removing elements if its length is not fixed.

\item \textbf{Structure.} Values with custom structure types are created by recursively generating their fields. When mutating custom structure values, \tool randomly selects a field in the structure and mutates it according to its type.

\item \textbf{Trivial Pointer.}  
Trivial pointers expose the layout of the pointer type (i.e., pointers to primitive data types and structure types). \tool mutates them by the following operations: 
(1) Set to a null pointer;
(2) Point to the location of an existing statement that holds the same pointer type;
(3) Point to a newly generated array whose element type is the same as the pointee;
(4) Point to the location of the value returned by a newly generated call statement. The call statement returns the same pointer type.
 
\item \textbf{Nontrivial Pointer.}  
Nontrivial pointers, such as opaque pointers, void pointers, and function pointers, cannot be mutated straightforwardly due to their unpredictable nature. Hence, \tool constructs them separately.
For the opaque pointers, \tool retrieves the API functions that initialize the pointers via either returning the pointers or filling them through reference. If void pointers have alias types, \tool treats them as opaque pointers as well. Otherwise, \tool attempts to cast an arbitrary-length byte array to the required void pointer to see if it works (detailed in \autoref{sec:intraapi}). 
For the function pointers, \tool synthesizes empty functions with the required signatures at compiling stage to allow the fuzzer to use their addresses as pointers.

\end{itemize}

Besides, byte arrays may contain data with their own encoding, which is not defined in header files. For example, \lstinline|cJSON_ParseWithOpt| in \autoref{fig:driver_example} parses a byte array with JSON formatting. For such values, \tool applies AFL's random mutations that are designed for byte arrays. 

\subsubsection{Input Minimization}
\label{sec:minimize}

Redundant statements and values slow down execution speed and increase search space during mutating, which makes fuzzing inefficient. 
To address this, \tool employs two steps to minimize the inputs:

\begin{enumerate}
\item [1.] Minimize inputs after mutation and refinement. \tool inspects the statement in the input backward, excluding the target call statement. If a statement is no longer referenced by other statements, \tool deletes it. For example, in \autoref{fig:workflow}, Line 0 in (c) is redundant after mutation and is thus removed by \tool in (d).
    
\item [2.] Minimize inputs that trigger new paths. \tool removes calls that have no impact on the execution path (detailed in \autoref{sec:interapi}), as well as redundant values in load statements. It tries to set the pointer values to null or shrink the length of arrays when possible. If the execution path remains unchanged, \tool retains the mutation in the input.
\end{enumerate}

\subsection{Constraint Learning}
\label{sec:constraint}
To correctly invoke APIs, DSL programs generated by \tool must satisfy both intra- and inter-API constraints. Intra-API constraints dictate that APIs must be invoked with appropriate arguments, while inter-API constraints specify the appropriate order in which APIs are invoked. \tool learns these constraints through libraries' runtime feedback rather than relying on external sources.

\subsubsection{Intra-API Constraint}
\label{sec:intraapi}
We propose six general intra-API constraints based on our observations of real-world libraries. 
 \begin{itemize}
    \item \textbf{Non-null Pointer(NON-NULL)}. APIs that do not check for null pointers can crash when invoked with null pointers.  It's often unclear whether this is a real bug, as some developers argue it is the user's responsibility to perform null checks. 
    \item \textbf{Valid File Resource(FILE)}. When an API function reads from or writes to a file, the file name provided as an argument must be valid. If the file name is randomly generated, the API call may terminate early, or it could mess up the disk if used as an output stream.
    \item \textbf{Specific Value(EQUAL)}. APIs that use a number in the arguments to designate the boundary of an array pointer may suffer from overflow errors if the number is incorrect.
    \item \textbf{Bounded Range(RANGE)}. APIs that access or allocate limited resources based on argument numbers may encounter resource exhaustion or overflow errors if the number is out of range.
    \item \textbf{Array Length(ARRAY-LEN)}. Some APIs assume that the arrays referenced by the pointers have sufficient elements rather than asking for arguments indicating boundaries. This may result in overflow errors if the arrays don't have enough elements.
    \item \textbf{Specific Type Cast(CAST)}. Due to missing layout information of the void type, developers have to generate objects with concrete types and cast their references to the void pointers.
\end{itemize}

\tool learns intra-API constraints throughout the entire fuzzing process, including both the pilot and evolution phases. 
In addition to inferring constraints for the arguments themselves, \tool recursively explores composite structures to infer constraints for any objects contained within, such as the pointees, fields in structs, and elements within arrays.

For \verb|void*| arguments without alias types, \tool assumes that their pointees do not contain any pointer. Therefore, they can be cast from a large enough random byte array, and \tool adds CAST constraints that treat them as \verb|char*| type. Conversely, in the case of \verb|void*| pointers with alias types, \tool supposes that they are opaque pointers and initializes them by invoking other APIs.

When new paths are explored by inputs, \tool checks to see if any file open function (e.g., \verb|fopen|) has been triggered, and compares the file name with the arguments used to invoke the API. If there is a match, a FILE constraint is created for the corresponding argument.

Furthermore, if an input triggers a new crash, \tool infers intra-API constraints by the following steps in order.
\begin{enumerate}
    \item If the call triggers a segmentation fault due to accessing a null pointer ($si\_addr$ is $0$ or close to $0$, where $si\_addr$ is the address of the faulting memory reference),  \tool locates each null pointer in the arguments, sets it to the address of a protected memory chunk, and runs this mutated program again. 
    If the program crashes again at the same program location (indicated by the RIP register) and triggers illegal access inside the protected memory chunk, it means the pointer is accessed without a null check in the API invocation. In that case, \tool adds a NON-NULL constraint for this pointer.
    \item If the crash is caused by accessing a canary appended right after an array ($si\_addr$ is in the range of canary), \tool tries to figure out whether there is a length or index of a variable-sized array in the arguments. Firstly, \tool locates which array has been overflowed. We denote the array's length as $N$. For each numerical value in the call's arguments, \tool attempts to set it as $N-1$, $N$, and $N+1$, respectively. If both $N$ and $N+1$ lead to a crash by accessing the canary, \tool adds a RANGE constraint to set the value within a range of $[0, N)$. If only $N+1$ makes a crash, an EQUAL constraint is added to set the value to be the same as the array's length. 
    \item If the above strategies fail to rectify illegal access of a canary, \tool attempts to pad the arrays in the arguments to a specific length $K$ (e.g., $64$) to see whether it resolves the crash. If so, an ARRAY-LEN constraint is added to ensure that this array is at least $K$ bytes.
    \item For other illegal access, if there are CAST constraints for the arguments, \tool tries to mutate the byte array pointed to by the arguments. If the illegal address varies with the mutated bytes, the \verb|void| pointer may be interpreted as a structure containing pointers. Thus, \tool removes the CAST constraints with \verb|char*| type. 
    \item If the inputs lead to a timeout or out of memory, \tool searches for large numerical values in the arguments and mutates them. If the execution becomes significantly faster or exits normally after setting the value to be small, \tool adds a RANGE constraint for the argument to limit its maximum value.
\end{enumerate}

These constraints are used to refine inputs after mutation, and inputs that violate the constraints are excluded from the seed set. As an example, in \autoref{fig:workflow}, \tool successfully infers the constraint that the value of \verb|len| should be the length of the array that \verb|name| refers to, and refines the value accordingly. 

\subsubsection{Inter-API Constraint}
\label{sec:interapi}

An API call can modify its referenced arguments or internal states in the program, which affects the behavior of its subsequent calls. These relationships between API calls are referred to as inter-API constraints. Compared to intra-API constraints, inter-API constraints are more difficult to identify and apply universally. Each library defines its own pattern to coordinate the API functions, making it hard to make abstractions. Instead of imposing concrete constraints, \tool preserves the inter-API constraints by saving the effective programs for later mutations.
To learn such constraints, \tool statically assembles library APIs by function signatures and then verifies their effectiveness dynamically through coverage feedback.

Initially, \tool statically infers naive constraints between API functions by analyzing their signatures. By comparing the types of their arguments and returns, we can deduce whether the two API functions are likely related. For any given pair of API functions $\mathbf{F_1}$ and $\mathbf{F_2}$, they tend to be related if their types overlap in the following ways.
\begin{itemize}
    \item Type $\mathbf{T}$ is the type of one of the arguments in $\mathbf{F_1}$, and it is the type of return value of $\mathbf{F_2}$.
    \item Both $\mathbf{F_1}$ and $\mathbf{F_2}$ use type $\mathbf{T}$ as their arguments.
    \item $\mathbf{F_1}$ uses an argument with type $\mathbf{T}$ while $\mathbf{F_2}$ uses a pointer $\mathbf{T}\ast$ that may modify the value it points to.
\end{itemize}
Moreover, if the identifiers of arguments in the latter two cases are identical, the two API functions are more likely to be related.

Two API functions have an effective relation if the former can help the latter to reach a new path. As shown in \autoref{alg:callgen}, \tool attempts to generate a large number of various API usages. The API calls may be composited in an interesting order after several iterations. In case \tool finds a certain sequence of API calls to be effective, the program is reserved for further mutation. By precisely controlling coverage tracking, as described in \autoref{sec:interpreter}, \tool learns the effective relations between API functions by the following steps.
\begin{enumerate}
\item When new calls are inserted before the target API call, only the coverage of the target call will be tracked.
\item If the insertions trigger a new path, \tool deletes the new calls one by one from the input and checks whether the coverage remains the same after each removal. If the coverage changes, \tool identifies the removed call as critical for reaching the new path. In other words, an efficient inter-API constraint between the two API functions is found in the program. 
\end{enumerate}

In addition to recognizing effective call sequences, \tool also identifies the effective arguments among sequences of API calls. Once a program triggers new paths, \tool checks whether the new paths were introduced by the mutation of certain arguments. 
To reuse effective arguments for further mutation, \tool caches all the statements that produce these effective arguments. These statements contain both concrete values and inter-API constraints.
During mutation, \tool replaces the original argument with an effective argument from the cache to introduce the appropriate context to the API calls.
\label{contextconstraint}
\section{Implementation}

We implemented \tool in 23164 lines of Rust code and 1285 lines of C/C++ code for instrumentation.

\subsection{Fuzzer}
\tool generates inputs with semantics extracted from C header files. Combining Rust's traits and macros, we implemented the process elegantly.
First, we adopted Rust bindgen~\cite{bindgen} to automatically generate Rust FFI bindings, including type definitions and function signatures, from library header files. For C++ libraries, \tool only accepts C-style API declarations since bindgen~\cite{bindgen_cpp} does not yet support features like templates.
Next, we defined \verb|mutate|, \verb|generate|, \verb|serialize|, \verb|deserialize| traits for the types in the bindings, which apply customized behavior to objects of each type for fuzzing. We manually implemented these traits for primitive types since they have common memory layouts in libraries. As for custom structures, we utilized Rust's \textit{procedural macros} to automatically implement these traits. 
\tool also synthesizes empty functions for function pointer types according to their function signatures. When invoked as callbacks, these functions do nothing other than return an object with zero-initialization.
Finally, we generated a table of object builders that \tool uses to call the trait implementations. These automatically generated codes are compiled and linked against the fuzzer.

To report the inputs that trigger bugs to developers for analysis, we have built a tool that translates DSL to C source code.

\subsection{Interpreter}

\tool instruments the library binaries using static binary rewriting with E9Patch~\cite{e9patch}.
We borrowed from E9AFL~\cite{e9afl} the code for branch tracking and improved it to be controllable and API-sensitive. Moreover, the instrumentation collects the compare instructions and hooks resources management functions (e.g., \verb|malloc|, \verb|free| and \verb|fopen|). 

When executing inputs, \tool's interpreter parses them based on the code that implements \verb|deserialize| traits. To invoke the APIs, \tool generates the code of the caller mapping table described in \autoref{sec:interpreter}, using Rust FFI bindings and procedural macros. This code is then compiled into the interpreter. Similar to AFL, the interpreter uses the techniques of fork servers to reduce the overhead of process initiation. It forks a new process once it receives a new input for interpretation.

During interpretation, \tool detects memory overflow by placing a page-size canary after the array values in a memory arena. This memory arena is implemented by \verb|mmap| continuous memory at a fixed address. In the arena, the address of the last byte of an array is aligned to a page's last byte. To detect overflow, \tool reserves a page size memory after the array and sets the page as unreadable and unwritable by \verb|mprotect|. 

\section{Evaluation}
In this section, we evaluated \tool on 11 widely-used real-world libraries. All of the selected libraries are commonly used by various applications and have been evaluated by OSS-Fuzz~\cite{serebryany2017oss}. Their developers have crafted several fuzz drivers to adapt fuzzing on those libraries using LibFuzzer. 
Among all the 11 libraries, \file{re2} is distinct in that it is developed and exported as a C++ library. For the sake of compatibility, \tool fuzzed \file{re2} through its C wrapper: \file{cre2}~\cite{cre2}. Additionally, all libraries included in our experiments were the latest versions available at the time of testing.
To demonstrate the effectiveness of \tool, the following research questions were answered:
\begin{itemize}
    \item \textbf{RQ1:} How effective is \tool in libraries?
    \item \textbf{RQ2:} Can \tool correctly infer API constraints?
    \item \textbf{RQ3:} Can \tool generate programs that are comparable with MCFs?
\end{itemize}

All of our experiments are conducted on a server with an Intel Xeon Platinum 8255C and 128 GB memory running 64-bit Ubuntu 20.04 LTS. In each experiment, we configured the tested fuzzer to be executed on only one core. 

\begin{table}[p] 
\vspace{-1em}
\setlength{\tabcolsep}{\tablength}
\setlength\extrarowheight{-2.5pt}
\caption{Comparison of coverage between \tool, Graphfuzz, FuzzGen, and MCFs on real world libraries.}

\resizebox{0.97\columnwidth}{!}{
\begin{tabular}{llrr}
\toprule
\textbf{Library} & \textbf{Fuzzer} & \textbf{Lines}  & \textbf{Branchs} \\
\midrule
\multirow{2}{*}{\file{cJSON}} & cjson\_read\_fuzzer & 952  (42.92\%)  & 473 (46.56\%)  \\
               \cmidrule{2-4} & FuzzGen             & 186  (8.39\%)   & 96 (9.45\%)  \\
               \cmidrule{2-4} & GraphFuzz           & 1346 (60.69\%)  & 612 (60.24\%)  \\
               \cmidrule{2-4} & \tool               & 1997 (90.04\%)  & 854 (84.06\%)  \\
\midrule
\multirow{4}{*}{\file{c-ares}} & parse\_reply\_fuzzer  & 1757 (23.94\%) & 744  (21.78\%)  \\
                               & create\_query\_fuzzer & 110  ( 1.50\%) &  47  ( 1.38\%)   \\
                               & Total of MCFs         & 1865 (25.41\%) & 790  (23.13\%)  \\
                \cmidrule{2-4} & GraphFuzz             & 2424 (33.02\%)  & 994 (29.10\%)  \\
                \cmidrule{2-4} & \tool                 & 5012 (67.06\%) & 2045 (59.03\%)  \\
\midrule
\multirow{2}{*}{\file{libpng}} & libpng\_read\_fuzzer & 5005 (28.67\%)  & 2041 (26.42\%)  \\
                \cmidrule{2-4} & GraphFuzz            & 1629 (9.33\%)   & 507 (6.56\%)  \\
                \cmidrule{2-4} & \tool                & 9610 (55.04\%)  & 3903 (50.53\%)  \\
\midrule
\multirow{2}{*}{\file{lcms}}   & IT8\_load\_fuzzer              &   641 ( 3.41\%)  &  271 (3.21\%)  \\
                               & overwrite\_transform\_fuzzer   &  2964 (15.77\%)  & 1271 (15.07\%)  \\
                               & transform\_fuzzer              &  4382 (23.32\%)  & 1757 (20.84\%)  \\
                               & Total of MCFs                  &  5357 (28.50\%)  & 2205 (26.15\%)  \\
                \cmidrule{2-4} & GraphFuzz                      & 2481 (13.20\%)   & 872 (10.34\%)  \\
                \cmidrule{2-4} & \tool                          &  9001 (47.89\%)  & 3135 (37.18\%)  \\
\midrule
\multirow{2}{*}{\file{libmagic}} & magic\_fuzzer & 3094 (30.87\%)  & 2043 (28.69\%)  \\
                \cmidrule{2-4} & GraphFuzz       & 3197 (31.90\%)   & 2003 (28.10\%)  \\
                \cmidrule{2-4} & \tool           & 4230 (45.26\%)  & 2634 (39.51\%)  \\
\midrule
\multirow{2}{*}{\file{libpcap}} & fuzz\_both     & 4301 (27.05\%)  & 2346 (32.40\%)  \\
                                & fuzz\_filter   & 5561 (34.97\%)  & 2901 (40.07\%)  \\
                                & fuzz\_pcap     &  974 ( 6.13\%)  &  365 ( 5.04\%)  \\
                                & Total of MCFs  & 6870 (43.21\%)  & 3479 (48.05\%)  \\
                 \cmidrule{2-4} & GraphFuzz       & 1736 (10.92\%)   & 881 (12.17\%)  \\
                \cmidrule{2-4} & \tool           & 7536 (47.40\%)  & 3669 (50.68\%)  \\
\midrule
\multirow{2}{*}{\file{zlib}} & checksum\_fuzzer         &  251 ( 5.21\%)  &   61 ( 2.12\%)  \\
                             & compress\_fuzzer         & 1880 (38.73\%)  &  904 (31.48\%)  \\
                             & example\_dict\_fuzzer    & 1901 (39.17\%)  &  952 (33.15\%)  \\
                             & example\_flush\_fuzzer   & 1631 (33.61\%)  &  776 (27.02\%)  \\
                             & example\_large\_fuzzer   & 1684 (34.70\%)  &  795 (27.68\%)  \\
                             & example\_small\_fuzzer   & 1429 (29.45\%)  &  758 (26.39\%)  \\
                             & minigzip\_fuzzer         & 2227 (45.89\%)  & 1071 (37.29\%)  \\
                             & uncompress2\_fuzzer      &  989 (20.37\%)  &  468 (16.30\%)  \\
                             & uncompress\_fuzzer       &  976 (20.11\%)  &  457 (15.91\%)  \\
                             & Total of MCFs            & 2976 (61.32\%)  & 1482 (51.60\%)  \\
             \cmidrule{2-4} & GraphFuzz       & 2602 (53.62\%)   & 1343 (46.76\%)  \\
              \cmidrule{2-4} & \tool                    & 3502 (72.16\%)  & 1914 (66.64\%)  \\
\midrule
\multirow{2}{*}{\file{re2}} & re2\_fuzzer & 6373 (67.85\%)  & 3367 (67.75\%)  \\
             \cmidrule{2-4} & GraphFuzz   & 5564 (59.23\%)  & 2843 (57.21\%)  \\
             \cmidrule{2-4} & \tool       & 6413 (68.27\%)  & 3299 (66.38\%)  \\

\midrule
\multirow{2}{*}{\file{sqlite3}} & oss\_fuzz   & 22582 (30.96\%)  & 9054 (25.99\%)  \\
                \cmidrule{2-4} & GraphFuzz    &  6261 ( 8.58\%)  & 2172 ( 6.23\%)  \\
                \cmidrule{2-4} & \tool        & 25356 (34.76\%)  & 9551 (27.41\%)  \\
\midrule
\multirow{2}{*}{\file{libvpx}} & vpx\_dec\_fuzzer\_vp8    & 3604 (10.44\%)   & 1138 (14.21\%)  \\
                               & vpx\_dec\_fuzzer\_vp9    & 15654 (45.34\%)  & 3475 (43.41\%)  \\
                               & Total of MCFs            & 18787 (54.42\%)  & 4463 (55.75\%)  \\
                \cmidrule{2-4} & FuzzGen                  & 15211 (44.06\%)  & 3367 (42.05\%)  \\
                \cmidrule{2-4} & GraphFuzz                & 15060 (43.62\%)  & 3251 (40.61\%)  \\
                \cmidrule{2-4} & \tool                    & 15641 (45.30\%)  & 3514 (43.89\%)  \\
\midrule
\multirow{2}{*}{\file{libaom}} & av1\_dec\_fuzzer   & 39762 (21.05\%)  & 9722 (17.73\%)  \\
                \cmidrule{2-4} & FuzzGen            & 32576 (17.25\%)  & 7757 (13.35\%)  \\
                \cmidrule{2-4} & GraphFuzz          & 30837 (16.33\%)  & 7327 (12.61\%)  \\
                \cmidrule{2-4} & \tool              & 36218 (19.18\%)  & 9228 (15.88\%)  \\
\bottomrule
\end{tabular}
}
\label{tbl:cov}
\vspace{-1em}
\end{table}

\subsection{RQ1: How effective is \tool in libraries? }
\label{sec:coverage}
We compared \tool with MCFs and other automatic solutions (i.e., FuzzGen and GraphFuzz) on 11 widely-used real-world libraries, using code coverage and bug finding as metrics.
For the MCFs, we selected fuzz drivers either crafted by the library developers or collected from the Google OSS-Fuzz project. 
The authors of FuzzGen provided the fuzz drivers for \file{libvpx} and \file{libaom}, which we used directly in our evaluation after rectifying the inaccuracies listed in \autoref{sec:wrong_drivers}. However, FuzzGen encountered errors during fuzz driver generation, except for \file{cJSON}. These errors involved violating assertions while determining data types of arguments or getting stuck at static backward slicing.  
GraphFuzz requires manually written schemas to synthesize fuzz drivers. We adopted the author-provided schema for \file{sqlite3} and wrote the others ourselves. 
In each individual experiment, we ran the fuzzers with a timeout of 24 hours. Afterward, we recompiled the tested library with LLVM's source-based code coverage feature~\cite{llvm-cov} enabled and re-ran the fuzzers with seed inputs to collect coverage information.
To reduce statistical errors, we repeated each experiment five times and reported the average results. Additionally, for libraries containing multiple MCFs, we ran each driver with LibFuzzer for 24 hours separately.

\subsubsection{Code Coverage}
\autoref{tbl:cov} shows the code coverage achieved by different fuzzers for each library. \tool outperformed other fuzzers on all libraries in both lines and branches coverage, except for \file{re2}, \file{libvpx}, and \file{libaom}. Even though 9 MCFs of \file{zlib} have been tested for a total of 216 hours with Libfuzzer, \tool achieved better results than the MCFs overall. Notably, \tool generated inputs that covered almost three times as much code as the MCFs' best effort for \file{c-ares}. On the other hand, while the MCFs for \file{re2} were sophisticated enough to cover most of the source codes, \tool managed to achieve a coverage level that is comparable to the MCFs. In the case of \file{libvpx} and \file{libaom},  MCFs covered more lines and branches than automatic solutions. This is because they can reach complicated internal states by decoding multiple frames in a loop, which is not possible for the automatic solutions that are loop-free. However, among the automatically generated fuzz drivers, \tool explores more lines and branches than FuzzGen and GraphFuzz.

To demonstrate that \tool is capable of fuzzing more APIs, we counted the number of unique APIs invoked by the valid programs generated by fuzzers. \autoref{tbl:api_cov} shows that, on average,  \tool successfully invoked 93.52\% of APIs, while MCFs, FuzzGen, and GraphFuzz only cover 18.58\%, 13.93\%, and 41.42\% APIs respectively.

\subsubsection{Bug Finding}\label{sec:bug_finding}


For the crashes triggered by \tool, we first eliminated any spurious crashes that violated the inter-API constraints learned by \tool and then removed the duplicated crashes with the same stack traces at the program crash point.
The remaining crashes were verified manually by inspecting the code, debugging the programs, and reading the official documents. \tool ultimately discovered 25 new bugs from 51 unique crashes, as shown in \autoref{tbl:crashes}. In contrast, none of the MCFs, FuzzGen, and GraphFuzz found a valid bug under the same 24 hours running.
Of the 51 unique crashes, 26 were identified as spurious crashes as they violated the constraints specified in the documentation or were rejected by library developers. The developers did not classify these crashes as bugs since they have their own criteria for valid API usage. The APIs must be used strictly within specified constraints, which may be mentioned in the documentation, or the program may crash unexpectedly. For example, an out-of-bound read error could occur if the \verb|zFunctionName| argument for \verb|sqlite3_create_function16| is not a UTF-16 string~\footnote{https://sqlite.org/forum/forumpost/7ace1408b.}.

We have reported all the discovered bugs to the corresponding communities. At the time of writing, 17 bugs have been confirmed by library developers and the rest are still under review. The details of the discovered bugs are listed in \autoref{tbl:bugs}. 
Null pointer dereference is the most common type of bug detected by \tool (11 out of 25). These bugs occurred when null pointers were produced or improperly initialized by library calls and accessed without null pointer checking within the code of API functions.
Buffer overflow is another common type of bug detected (8 out of 25). Six of those overflows were detected by \tool's canary when reading or writing out of bounds of arrays allocated as arguments by \tool. The two remaining bugs involved accessing either uninitialized memory or the addresses of internally allocated arrays in the API functions beyond their bounds.
Apart from these, \tool identified three division by zero errors, an uncontrolled format string bug, a double-free bug, and an infinite loop, which are analyzed in detail in \autoref{sec:case_study}.

Most of the buggy APIs discovered by \tool were not covered by MCFs, FuzzGen, or GraphFuzz. This is not surprising as the fuzz drivers for these libraries have been extensively utilized for continuous fuzzing over an extended period. 
To ensure a fair comparison with MCFs, GraphFuzz, and FuzzGen, we additionally evaluated \tool and these fuzzers on previous versions of the 11 libraries. 
As shown in \autoref{tbl:crashes_old}, \tool outperformed MCFs, GraphFuzz, and FuzzGen by detecting 28 unique bugs in previous versions of the 11 libraries. Moreover, \tool was able to identify 7 out of the 14 bugs that other fuzzers had detected. Although \tool missed certain bugs reported by other fuzzers due to its inefficient mutation power on those predefined API calls, it prioritized mutating API usages that are more likely to trigger bugs or explore new states via its seed selection, which includes any API usage within the libraries. Consequently, \tool discovered a greater number of bugs overall.


\begin{table}[h]
\caption{Unique crashes reported by \tool.}
\setlength{\tabcolsep}{\tablength}
\resizebox{0.97\columnwidth}{!}{

\begin{threeparttable}
\begin{tabular}{p{0.14\columnwidth}>{\centering}p{0.15\columnwidth}>{\centering}p{0.19\columnwidth}>{\centering}p{0.08\columnwidth}>{\centering}p{0.08\columnwidth}>{\centering}p{0.08\columnwidth}>{\centering}p{0.08\columnwidth}cp{0.2\columnwidth}}
\toprule
\multirow{2}{*}{\textbf{Library}}     &\multirow{2}{*}{\textbf{Version}}  &\multirow{2}{*}{\textbf{\#Programs}}     &\multicolumn{4}{c}{\textbf{Crashes}}                     & \multirow{2}{*}{\textbf{Accuracy}}\\
\cmidrule{4-7}
                                      &                                   &                                   &\textbf{\#UC} & \textbf{\#S} & \textbf{\#B} & \textbf{\#C} &  \\
\midrule
\file{cJSON}           & 1.7.15          & 2,972                & 3              & 1            & 2            & 2      & 66.67\%      \\
\file{c-ares}          & 1.18.1          & 3,192                & 2              & 0            & 2            & 2      & 100\%       \\
\file{libpng}          & 1.6.37          & 5,612                & 8              & 4            & 4            & 1      & 50\%       \\
\file{lcms}            & 2.13.1          & 2,660                & 13             & 8            & 5            & 5      & 38.46\%    \\
\file{libmagic}        & FILE5\_42       & 1,662                & 0              & 0            & 0            & 0      & -          \\
\file{libpcap}         & 1.10.1          & 2,249                & 5              & 2            & 3            & 3      & 60\%    \\
\file{zlib}            & 1.2.12          & 5,598                & 4              & 1            & 3            & 3      & 75\%       \\
\file{re2}             & 0.4.0           & 27,355               & 4              & 2            & 2            & 0      & 50\%    \\
\file{sqlite3}         & 3.38.5          & 10,356               & 11             & 7            & 4            & 1      & 36.36\%    \\
\file{libvpx}          & 1.11.0          & 22,282               & 0              & 0            & 0            & 0      & -          \\
\file{libaom}          & 3.5.0           & 19,654               & 1              & 1            & 0            & 0      & -          \\

\midrule
\textbf{Total}        & -            & \textbf{103,592}      & \textbf{51}  & \textbf{26}  & \textbf{25}   & \textbf{17}  & \textbf{49.02\%}   \\
\bottomrule
\end{tabular}
\small
\begin{tablenotes}
\item
\textbf{UC} = Total unique crashes; \textbf{S} = Spurious crashes caused by incorrect API usage; \textbf{B} = Valid bugs that identified by manually review; \textbf{C} = Confirmed bugs after reported to library developers. 
\end{tablenotes}
\end{threeparttable}
}
\label{tbl:crashes}
\end{table}

\begin{table*}[ht]
\caption{Verified bugs found by \tool.}
\setlength{\tabcolsep}{\tablength}
\resizebox{0.97\textwidth}{!}{
\begin{threeparttable}
\begin{tabular}{llllllllll}
\toprule
\textbf{ID} & \textbf{Library} & \textbf{Location} & \textbf{Buggy function} & \textbf{Bug Type}  & \textbf{Status} & \textbf{Commit ID} & \textbf{MCF} & \textbf{FZ} & \textbf{GF}\\
\midrule[0.8pt]
1.  & cJSON   & cJSON.c:2209 & cJSON\_DetachItemViaPointer  & Null pointer dereference  & Confirmed   & 722(P) & \XSolid & \XSolid & \XSolid  \\
2.  & cJSON   & cJSON.c:2326 & cJSON\_ReplaceItemViaPointer & Null pointer dereference  & Fixed       & 726(P) & \XSolid & \XSolid & \XSolid  \\
3.  & c-ares  & ares\_init.c:2254  & ares\_set\_sortlist    & Buffer overflow           & Fixed       & 496(S) & \XSolid & -       & \XSolid  \\
4.  & c-ares  & ares\_init.c:2247  & ares\_set\_sortlist    & Buffer overflow           & Fixed       & 496(S) & \XSolid & -       & \XSolid  \\
5.  & libpng  & pngerror.c:229     & png\_warning           & Buffer overflow           & Confirmed   & 453(S) & \XSolid & -       & \XSolid  \\
6.  & libpng  & pngwrite.c:1976   & png\_image\_write\_to\_file  & Division by zero  & Reported   & 489(S) & \XSolid & -       & \XSolid  \\
7.  & libpng  & pngwrite.c:1976   & png\_image\_write\_to\_memory  & Division by zero  & Reported   & 489(S) & \XSolid & -       & \XSolid  \\
8.  & libpng  & pngwrite.c:1976   & png\_image\_write\_to\_stdio  & Division by zero  & Reported   & 489(S) & \XSolid & -       & \XSolid  \\
9. & zlib    & deflate.c:401     & gzsetparams            & Null pointer dereference  & Fixed       & 761(P) & \XSolid & -       & \XSolid  \\
10. & zlib    & gzread.c:517    & gzungetc              & Buffer overflow  & Fixed    & 837(S) & \XSolid & -       & \XSolid  \\
11. & zlib    & gzwrite.c:136   & gzflush               & Infinite loop   & Fixed    & 840(S) & \XSolid & -       & \XSolid  \\
12. & re2     & cre2.cpp:195    & cre2\_find\_named\_capturing\_groups & Null pointer dereference  & Reported  & 30(S) & \XSolid & - & \XSolid  \\
13. & re2     & cre2.cpp:725      & cre2\_set\_match       & Null pointer dereference  & Reported    & 29(S)  & \XSolid & -       & \XSolid  \\
14. & sqlite3 & sqlite3.c:30121    & sqlite3\_overload\_function & Uncontrolled format string & Fixed  & bbbbb66b6b(T) & \XSolid & -   & \XSolid  \\
15. & sqlite3 & sqlite3.c:79786    & sqlite3\_value\_bytes  & Null pointer dereference  & Reported  & 0218d74c47(T) & \XSolid & -   & \XSolid  \\
16. & sqlite3 & sqlite3.c:79786    & sqlite3\_value\_bytes16 & Null pointer dereference & Reported  & 0218d74c47(T) & \XSolid & -   & \XSolid  \\
17. & sqlite3 & sqlite3.c:85266    & sqlite3\_value\_subtype & Null pointer dereference & Reported  & 0218d74c47(T) & \XSolid & -   & \XSolid  \\
18. & lcms    & cmsio1.c:857       & cmsIsCLUT           & Buffer overflow       & Fixed           & 350(S)   & \XSolid & -       & \XSolid  \\
19. & lcms   & msgamma.c:852       & cmsBuildTabulatedToneCurveFloat   & Buffer overflow   & Fixed   & 351(S)  & \XSolid & -       & \XSolid  \\
20. & lcms   & cmsnamed.c:760      & cmsGetPostScriptCRD    & Null pointer dereference   & Fixed    & 353(S)  & \XSolid & -       & \XSolid  \\
21. & lcms   & cmslut.c:416        & cmsStageAllocMatrix    & Buffer overflow        & Fixed     & 354(S)   & \XSolid & -       & \XSolid  \\
22. & lcms   & cmscgats.c:1928     & cmsIT8SaveToMem        & Buffer overflow       & Fixed     & 355(S)   & \XSolid & -       & \XSolid  \\
23. & libpcap  & pcap.c:2946       & pcap\_breakloop        & Null pointer dereference   & Fixed   & 1147(P)   & \XSolid & -       & \XSolid  \\
24. & libpcap  & pcap.c:493        & pcap\_can\_set\_rfmon  & Null pointer dereference   & Fixed   & 1147(P)   & \XSolid & -       & \XSolid  \\
25. & libpcap  & pcap-linux.c:835  & pcap\_activate         & Double Free      & Fixed       & 1098(S)   & \XSolid & -       & \XSolid  \\
\bottomrule
\end{tabular}
\small
\begin{tablenotes}
\item
\textbf{Location} : The source file location of the crash point.
\textbf{Buggy function} : The API function that triggers the crash.
\item
\textbf{Commit ID} : The bug trace id that was committed to community developers. \textbf{P}=Pull request number, \textbf{S}=Issue number, \textbf{T}=Bug forum id.
\item
\textbf{MCF, FZ, GF} : Whether the bug could be discovered by MCFs, FuzzGen (FZ), and GraphFuzz (GF) under the same library version.
\end{tablenotes}
\end{threeparttable}
}
\label{tbl:bugs}
\end{table*}

\begin{table}[h]
\caption{Comparison with other fuzzers on previous versions of libraries.}
\setlength{\tabcolsep}{\tablength}
\resizebox{0.97\columnwidth}{!}{
\begin{threeparttable}
\begin{tabular}{ccccc|c|c}
\toprule
\textbf{Library} & \textbf{Version} & \textbf{MCF}  & \textbf{GraphFuzz}  & \textbf{FuzzGen} & \textbf{MCF $\cup$ GF } & \textbf{Hopper} \\ 
\hline
\file{cJSON}          & 1.7.0                & 1                    & 0                    & 0         & 1           & 4(1)    \\
\file{c-ares}         & 1.16.0               & 1                    & 0                    & -         & 1           & 4(1)    \\
\file{libpng}         & 1.6.32               & 0                    & 0                    & -         & 0           & 4       \\
\file{lcms}           & 2.8                  & 2                    & 1                    & -         & 2           & 6(1)    \\
\file{libmagic}       & FILE5\_25            & 1                    & 1                    & -         & 1           & 1(1)    \\
\file{libpcap}        & 1.9.0                & 4                    & 0                    & -         & 4           & 3(3)    \\
\file{zlib}           & 1.2.11               & 0                    & 0                    & -         & 0           & 1       \\
\file{sqlite3}        & 3.22.0               & 1                    & 0                    & -         & 1           & 3       \\
\file{libaom}         & 1.0.0                & 3                    & 1                    & 0         & 4           & 2       \\ 
\midrule
\textbf{Total}        & -                    & 13                   & 3                    & 0         & 14          & 28(7)   \\ 
\bottomrule
\end{tabular}
\small
\begin{tablenotes}
\item
Numbers in brackets show the number of bugs that were reported by multiple fuzzers including \tool. 
\end{tablenotes}
\end{threeparttable}
}
\label{tbl:crashes_old}
\end{table}

\smallskip
\begin{mybox}[boxsep=0pt,
 boxrule=1pt,
 left=4pt,
 right=4pt,
 top=3pt,
 bottom=3pt,
 ]
~\textbf{Answer to RQ1:} 
\tool outperformed MCFs and other library fuzzing approaches in terms of both code coverage and bug finding. Specifically, \tool detects 25 new bugs and 17 of them have been confirmed.
\end{mybox}

\subsection{RQ2: Can \tool correctly infer API constraints? }\label{sec:eval_constraint}

One of the key insights that allows \tool to generate correct and efficient API usages is its ability to learn intra- and inter-API constraints.
To evaluate the benefits of these constraints, we compared \tool with and without constraints on the 11 libraries. 
As shown in \autoref{tbl:constraint-analysis}, \tool with inferred constraints explores more lines and branches across all the libraries. 
However, without intra-API constraints, code coverage is substantially lower in \file{libpng}, \file{libmagic}, \file{libvpx}, and \file{libaom}. This is because certain API functions that many other APIs depend on could not be properly called. For instance, in \file{libpng}, \verb|png_init_io| initializes a file input for a handler that is subsequently read by other APIs for processing. If the fuzzer randomly generates a string as the file's name without a FILE constraint, \verb|png_init_io| will return an invalid handler, causing all subsequent calls to terminate at a shallow state.
As incorrect API usages often lead to program crashes, we also measured the success rate of execution for the generated inputs. Notably, after imposing intra-API constraints, \file{cJSON}, \file{c-ares}, \file{libvpx}, and \file{libaom} achieve almost 100\% success rate of execution. In \file{re2}, the success rate improves significantly after \tool learns not to generate null pointers as arguments.
On the other hand, in the absence of inter-API constraints, \tool explores API usage in a less efficient way by blindly combining API functions. 
This leads to poorer performance in terms of both line and branch coverage, particularly in libraries such as \file{libpng}, \file{lcms}, and \file{sqlite3}, which have a rich set of APIs available and therefore a large number of possible combinations to try out. To accelerate the search process, \tool leverages inter-API constraints to pinpoint and reuse effective API usages.

Furthermore, we counted the number of intra-API constraints learned from the 11 libraries and analyzed their correctness. In total, \tool learned 973 intra-API constraints. Among them, NON-NULL constraints were the most commonly inferred, as 609 out of the 973 constraints ( 62.01\%) pertained to pointer dereferences  without null checks in the API code.
Additionally, \tool was able to infer 17 FILE constraints, 147 EQAUL constraints, 35 RANGE constraints, 110 CAST constraints, and 55 ARRAY-LEN constraints across these libraries. The overall precision and recall of the learned constraints are 96.51\% and 97.61\%, respectively.
Most of the false-positive constraints learned by \tool are actually approximations of the ground truth constraints and work well in most cases. These constraints, although not completely correct, do help to reduce the invalid search space and prevent spurious crashes. For example, in \lstinline|cmsStageAllocCLut16bitGranular|, setting the fourth argument \verb|outputChan| to be the length of the fifth argument \verb|table|, while not entirely accurate, worked effectively in most cases.
Regarding the recall of the learned constraints, we identified false-negative constraints by analyzing the spurious crashes reported by \tool since collecting the entire ground truth is extremely labor-consuming. Moreover, most of these missed constraints are specific to a particular library and hard to describe in a general way. Typically, a spurious crash indicates a violation of certain unlearned constraints. As an example from \file{sqlite3} described in \autoref{sec:bug_finding}, \tool cannot ensure that the generated buffers are strictly UTF-16 encoded, which leads to a spurious crash. 

\begin{table*}[ht]
\caption{Impact of inter- and intra-API constraints on \tool's efficiency.}
\setlength{\tabcolsep}{\tablength}
\resizebox{\textwidth}{!}{

\begin{threeparttable}
\begin{tabular}{l|cccccc|cccc|cc|ccc|ccc}
\toprule
\multicolumn{1}{c}{\multirow{2}{*}{\textbf{Library}}} & \multicolumn{10}{|c|}{\textbf{Learned intra-API Constraints}} & \multicolumn{2}{c|}{\textbf{Success Rate}} & \multicolumn{3}{c|}{\textbf{Line Coverage}} & \multicolumn{3}{c}{\textbf{Branch Coverage}} \\ \cmidrule(lr){2-19}  & \textbf{\#N} & \textbf{\#F} & \textbf{\#E} & \textbf{\#R} & \textbf{\#C} & \textbf{\#A} & \textbf{TTL} & \textbf{\#FN} & \textbf{PRC} & \textbf{RCL} &  \textbf{w/o Intra}  &  \textbf{\tool} &  \textbf{w/o Intra} &  \textbf{w/o Inter} & \textbf{\tool} &  \textbf{w/o Intra} &  \textbf{w/o Inter} & \textbf{\tool} \\ 
\midrule
\file{cJSON}  & 15  & 0 & 10    & 4  & 1   & 0   & 30   & 1  & 100\%  & 96.77\% & 99.4291\% & 99.9999\% & 80.93\% & 88.86\% & 90.04\% & 75.00\% & 82.48\% & 84.06\% \\
\file{c-ares} & 87  & 0 & 26(1) & 2  & 16(1) & 2 & 133(2) & 0 & 98.50\% & 100\% & 91.4087\%  & 99.9999\% & 47.78\% & 65.31\% & 67.06\% & 43.91\% & 56.90\% & 59.63\% \\
\file{libpng} & 34(1) & 2 & 20 & 5(1) & 14 & 16(6) & 91(8) & 1 & 91.21\% & 98.81\% & 98.8973\% & 99.9917\% & 12.03\% & 39.05\% & 55.04\% & 7.39\% & 36.46\% & 50.53\% \\
\file{lcms}  & 233 & 6 & 25(6) & 1(1)  & 25(3) & 13(3) & 303(13) & 8 & 95.71\% & 97.32\% & 95.3159\%  & 99.2425\% & 46.65\% & 28.37\% & 47.89\% & 34.54\% & 21.05\% & 37.18\% \\
\file{libmagic} & 3 & 1 & 2(1)  & 0 & 3  & 1 & 10(1) & 0  & 90.00\%   & 100\%   & 99.4833\%  & 99.9935\% & 13.16\% & 42.20\% & 45.26\% & 9.96\% & 36.08\% & 39.51\% \\
\file{libpcap} & 59  & 4 & 1(1) & 4  & 1 & 4(2) & 73(3) & 2 & 95.89\%  & 97.22\% & 94.4091\%  & 99.9797\% & 40.92\% & 40.89\% & 47.40\% & 43.56\% & 43.14\% & 50.68\% \\
\file{zlib}  & 6  & 1 & 20 & 3  & 6 & 6 & 42 & 1  & 100\%  & 97.67\%  & 97.0134\%  & 99.9865\% & 52.94\% & 59.28\% & 72.16\% & 44.46\% & 48.96\% & 66.64\% \\
\file{re2} & 93 & 0 & 26  & 0  & 0 & 4 & 123 & 2 & 100\% & 98.40\% & 86.1034\%  & 99.9427\% & 58.21\% & 63.72\% & 68.27\% & 54.81\% & 61.91\% & 66.38\% \\
\file{sqlite3} & 72 & 3 & 8 & 10(2) & 38(2) & 9(1) & 140(5) & 7 & 96.43\% & 95.07\% & 99.3814\%  & 99.5527\%  & 30.52\% & 29.44\% & 34.76\% & 23.72\% & 23.34\% & 27.41\% \\
\file{libvpx} & 3 & 0 & 2 & 1(1)  & 4 & 0 & 10(1) & 0 & 90.00\% & 100\% & 98.8642\%  & 99.9999\% & 1.14\% & 42.51\% & 45.30\% & 3.35\% & 42.61\% & 43.89\% \\
\file{libaom} & 4 & 0 & 7 & 5(1)  & 2 & 0 & 18(1) & 1 & 94.44\% & 94.44\% & 94.2413\% & 99.9999\% & 1.73\% & 16.67\% & 19.18\% & 3.12\% & 13.32\% & 15.88\% \\
\midrule
\textbf{Total}   & \textbf{609(1)} & \textbf{17}& \textbf{147(9)} & \textbf{35(6)}& \textbf{110(6)}& \textbf{55(12)}& \textbf{973(34)}  & \textbf{23} & \textbf{96.51\%} & \textbf{97.61\%}  & \textbf{90.4769\%} & \textbf{99.8808\%} & \textbf{35.09\%} & \textbf{46.94\%} & \textbf{53.85\%}  & \textbf{31.26\%} & \textbf{42.39} & \textbf{49.25\%} \\

\bottomrule
\end{tabular}
\begin{tablenotes}
\item
\textbf{N} = NON-NULL; \textbf{F} = FILE; \textbf{E} = EQUAL; \textbf{R} = RANGE; \textbf{C} = CAST; \textbf{A} = ARRAY-LEN; \textbf{TTL} = Total number; \textbf{FN} = False negative; \textbf{PRC} = Precision; \textbf{RCL} = Recall.
\item Numbers in brackets show the number of false positive constraints.
\end{tablenotes}
\end{threeparttable}
}
\label{tbl:constraint-analysis}
\end{table*}

\smallskip
\begin{mybox}[boxsep=0pt,
 boxrule=1pt,
 left=4pt,
 right=4pt,
 top=3pt,
 bottom=3pt,
 ]
    ~ \textbf{Answer to RQ2:}
    \tool has the ability to learn intra-API constraints with high precision (96.51\%) and recall (97.61\%), while also being capable of speeding up the search process during fuzzing through the use of inter-API constraints.
\end{mybox}

\subsection{RQ3: Can \tool generate programs that are comparable with MCFs? }\label{sec:case_study}

To demonstrate the ability of \tool for generating high-quality fuzz drivers, in \autoref{fig:eva_case}, we select two representative programs generated by \tool in our experiment as case studies.
\begin{figure}[t]
\begin{subfigure}[b]{1\linewidth}
\caption{An uncontrolled format string bug found in \file{sqlite3}.} \label{fig:case_sqlite}
\begin{lstlisting}[language=Hopper, numbers=none, xleftmargin=0.3em]
<0> load sqlite3 * = null //db
<1> load Vec<char> = vec(2)[96, 0] //file_buf
<2> file option <1> //filename
<3> load sqlite3 ** = &<0> //ppDb
<4> call relative: sqlite3_open ? (<2>, <3>) 
<5> assert non_null(<0>
<6> load Vec<char> = vec(22)["JSFuMPRKm/RVAABApl6+vy6ib8NjAA=="] //zFuncName, decoded as "%!n0JU@^.o"
<7> load char* = &<6>
<8> load i32 = 55767322 //nArg
<9> call target: sqlite3_overload_function ? (<0>,<7>,<8>)
\end{lstlisting}
\end{subfigure}


\begin{subfigure}[b]{1\linewidth}
\caption{A double free bug found in \file{libpcap}.}\label{fig:case_libpcap}
\begin{lstlisting}[language=Hopper, numbers=none, xleftmargin=0.3em]
 <0> load char * = null //device
 <1> load char * = null //errbuf
 <2> call pcap_create ? (<0>, <1>) //p
 <3> assert non_null(<2>)
 <4> load pcap_t * = <2> //p
 <5> load int = 128 //buffer_size
 <6> call relative: pcap_set_buffer_size ? (<4>, <5>)
 <7> load int = 15 //snaplen
 <8> call relative: pcap_set_snaplen ? (<4>, <7>)
 <9> load int = 1833782204 //immediate
<10> call relative: pcap_set_immediate_mode ? (<4>, <9>)
<11> call target: pcap_activate ? (<4>)
\end{lstlisting}
\end{subfigure}
\caption{Example programs that trigger bugs in \file{sqlite3} and \file{libpcap} respectively.}
\label{fig:eva_case}
\end{figure}

\textbf{Case 1.}
\autoref{fig:case_sqlite} shows an uncontrolled format string bug found in \file{sqlite3} (ID 14 in \autoref{tbl:bugs}). In \func{sqlite3\_overload\_function}, \func{zFuncName} is passed to \func{sqlite3\_mprintf} to clone a string, but at worst, \func{sqlite3\_mprintf}  would perform formatting if the string contains specifiers. 
\tool crafted this program successfully through the following steps. In the beginning, it generated a simple program that passes a null \func{sqlite3} pointer and a random \func{zFuncName} string to \func{sqlite3\_overload\_function}.  However, since the API function requires a non-null \func{sqlite3}, \tool used intra-API constraints to initialize the pointer by invoking \func{sqlite3\_open}. This made the program reach the code that calls \func{sqlite3\_mprintf}. Shortly thereafter, \tool randomly mutates the strings for \func{zFuncName} and fortunately produces a '\%' symbol in the string.
This case indicates that \tool is able to synthesize valid code to invoke an API function with intra-API constraints and find bugs using type-aware mutation for arguments.

\textbf{Case 2.}  \autoref{fig:case_libpcap} shows a double-free bug found in \file{libpcap} (ID 25 in \autoref{tbl:bugs}). 
This bug occurs when \func{pcap\_activate} fails to enable memory-mapped capture due to \func{mmap} with zero length under the immediate mode. In such a case, a memory location would be released twice in the error handling code. The length is computed from the combination of \func{buffer\_size} and \func{snaplen}.
Even worse, the \func{pcap\_activate} is a commonly used API function for activating a packet capture handle rather than those rarely-used ones.
\file{Tcpdump}, as a popular Linux application, is able to invoke those \file{libpcap} APIs sequentially by a simple command: \lstinline|tcpdump -i any -B 1 -s 15 --immediate-mode|.
By utilizing inter-API constraints, \tool carefully crafted a sequence of API calls that triggered the bug. Specifically, it inferred that \func{pcap\_set\_buffer\_size}, \func{pcap\_set\_snaplen}, and \func{pcap\_set\_immediate\_mode} would modify \func{pcap\_t} pointer to change the behavior of \func{pcap\_activate}. Then, \tool mutated the integers fed into the API calls and triggered the bug finally.
The case demonstrates that \tool can explore different API usages by learning inter-API constraints, which traditional fuzz testing tools typically ignore.

\smallskip
\begin{mybox}[boxsep=0pt,
 boxrule=1pt,
 left=4pt,
 right=4pt,
 top=3pt,
 bottom=3pt,
 ]
    ~ \textbf{Answer to RQ3:}
With grammar-aware input fuzzing, \tool synthesizes programs that satisfy both intra- and inter-API constraints. As compared to MCFs, the programs generated by \tool can explore a much broader range of API usages. 
\end{mybox}

\section{Discussions}
\subsection{Multiple-dimensional Search Space in Input Generation}\label{sec:search_space}

Traditional library fuzzing generates a byte array as input and leaves the job of constructing arguments to fuzz drivers. 
In contrast, in \tool, the search space for input generation is multidimensional as it involves both API functions and arguments, which poses a significant challenge. 
Furthermore, each argument has its own encoding format and requires specific mutation strategies.
Although \tool mitigates this issue by implementing novel techniques such as constraint learning and type-aware mutation, there is still much room for improvement.

\subsection{Compatibility with C++ Libraries}

Currently, \tool only supports fuzzing libraries with C-style header files. The use of templates in C++ headers delays the compilation of template functions to the time of instantiation by the users, thus making it challenging for \tool to generate callers of C++ functions and their arguments. Moreover, it is non-trivial to decide the concrete types of template parameters to instantiate the templates. To enable compatibility with C++, a more generalized implementation of generation and mutation is required. We plan to address this in future work.

\subsection{False Positive Crashes}
Although \tool is highly effective at inferring common constraints to filter out most spurious crashes, the remaining crashes may still be false positives since the APIs need to be used in specific ways. Learning these constraints through dynamic feedback can be challenging as they have no universal criteria, as discussed in \autoref{sec:eval_constraint}.
However, during fuzzing, \tool will no longer generate input for APIs that have failed to learn constraints and have a high probability of crashing spuriously.
To make \tool more practical and user-friendly, we plan to add warnings for users about unlearned constraints and provide a convenient way for them to add these custom constraints themselves.

\section{Related work}

\subsection{API Fuzzing}
APIs have long been favored as fuzzing targets, given their role in allowing code modules to interact with others. 
For example, RESTler generates requests for RESTful APIs using a grammar automatically inferred from Swagger specification and employs coarse-grained feedback from service responses to guide mutations to reach deeper service states~\cite{atlidakis2019restler}. Similarly, Pythia~\cite{atlidakis2020pythia} and MINER~\cite{miner2023} fuzz REST APIs using coverage-guided feedback and machine learning models.
In the realm of kernel fuzzers, Syzkaller~\cite{syzkaller} is a coverage-guided fuzzer that generates system call sequences using system API descriptions. To further improve the efficiency and effectiveness of kernel fuzzing, MoonShine~\cite{pailoor2018moonshine} generates seed inputs by extracting system call traces from the execution of real programs, while Healer~\cite{sun2021healer} uses dynamic analysis to learn the relationships between system calls and uses that knowledge to guide input generation.

Besides the adoption of fuzzing in REST services and system APIs, OSS-Fuzz~\cite{serebryany2017oss} gathers hundreds of manually written fuzzers to continuously fuzz APIs of open source libraries.  In recent years, there has also been a trend towards generating fuzz drivers for library APIs automatically~\cite{babic2019fudge, ispoglou2020fuzzgen, zhang2021apicraft, jung2021winnie, green2022graphfuzz}. FUDGE~\cite{babic2019fudge} and FuzzGen~\cite{ispoglou2020fuzzgen} extract the code of API usage from practical code to create fuzz drivers. APICRAFT~\cite{zhang2021apicraft} and WINNIE~\cite{jung2021winnie} record the API call sequences from the execution trace of existing consumer programs and combine them to generate fuzz drivers. GraphFuzz~\cite{green2022graphfuzz} views fuzz drivers as dataflow graphs and performs graph-based mutation with a manually specified schema. UTOPIA~\cite{jeong2022utopia} statically analyzes the library to identify attributes of API arguments and automatically synthesize valid fuzz drivers from existing unit tests.
In contrast to generating fuzz drivers in limited domains, \tool is a fuzzer for library APIs without any specific domain knowledge, including any knowledge from practical examples and schema specifications. \tool searches the combinations of APIs and arguments in a large space and learns the features of high-quality inputs simultaneously through runtime feedback.

\subsection{Grammar-aware Fuzzing}
Grammar-aware fuzzers leverage grammar to generate structured inputs that bypass syntax checking at the beginning of the program execution~\cite{holler2012fuzzing,veggalam2016ifuzzer,lee2020montage, zhong2020squirrel, wang2019superion,aschermann2019nautilus, sqlsmith}.
For example, LangFuzz~\cite{holler2012fuzzing} and IFuzzer~\cite{veggalam2016ifuzzer} utilize the syntax of JavaScript language to fuzz Javascript interpreters, while SQLsmith~\cite{sqlsmith} and SQUIRREL~\cite{zhong2020squirrel} generate SQL queries for testing DBMSs based on SQL grammar. 
Superion~\cite{wang2019superion} and NAUTILUS~\cite{aschermann2019nautilus} improve grammar-aware fuzzing by combining code coverage guidance. These tools manipulate input as an AST and mutate it according to coverage feedback.

As writing grammar rules require much human effort, some fuzzers try to automatically learn the grammar~\cite{wang2017skyfire,godefroid2017learn,blazytko2019grimoire,you2019profuzzer}.
Skyfire~\cite{wang2017skyfire} uses probabilistic modeling to learn grammar from inputs, while Learn\&Fuzz~\cite{godefroid2017learn} employs a recurrent neural network model. Nevertheless, the accuracy of the learned grammar is dependent on the quality of the corpus provided. Alternatively, GRIMOIRE~\cite{blazytko2019grimoire}  automatically infers the structural properties of input language based on code coverage feedback. Similarly, Profuzzer~\cite{you2019profuzzer} probes the types of input bytes through per-byte mutations.

General grammar-based fuzzers are intended to test parsers or interpreters by exploring all feasible combinations based on input grammar. However, \tool operates differently by focusing on generating various effective API calls rather than exhaustively exploring the input language's grammar.

\section{Conclusion}
In this paper, we present \tool, a novel fuzzer that aims to fuzz libraries without any domain knowledge required in crafting fuzz drivers.
\tool links the libraries under test against an interpreter, which takes DSL programs as input and drives libraries to perform requested fuzzing behavior. To generates effective API calls in the format of DSL, \tool learns intra- and inter-API constraints in the libraries and mutates the inputs with grammar awareness.
We evaluated the effectiveness of \tool on 11 real-world libraries. \tool outperformed MCFs and the other automatic solutions in both code coverage and bug finding. Specifically, \tool found 25 new bugs that others could not find.
Our experimental results demonstrate that \tool effectively explores a vast range of API usages for library fuzzing out of the box.

\section{Acknowledgment}

This material is based upon work supported by the National Science Foundation under Grant No.\ 1801751 and 1956364.

\printbibliography

@Online{afl,
  title = {American Fuzzy Lop},
  url = {http://lcamtuf.coredump.cx/afl/},
  author = {Michal Zalewski},
  %year  =   2017,
}

@online{libfuzzer,
  title = {{LibFuzzer} – a library for coverage-guided fuzz testing},
  url = {https://llvm.org/docs/LibFuzzer.html}
}

@online{llvm-cov,
 title = {Source-based Code Coverage},
 url = {https://clang.llvm.org/docs/SourceBasedCodeCoverage.html}
}

@online{syzkaller,
  title = {{syzkaller - kernel fuzzer}},
  url = {https://github.com/google/syzkaller}
}

@online{cre2,
  title = {{cre2}},
  url = {https://github.com/marcomaggi/cre2/}
}

@InProceedings{angora,
    author = 	 {Peng Chen and Hao Chen},
    title = 	 {Angora: efficient fuzzing by principled search},
    booktitle = {IEEE Symposium on Security and Privacy (S\&P)},
    address = {San Francisco, CA},
    year = {2018},
    month = {5},
}

@InProceedings{matryoshka,
    author = {Peng Chen and Jianzhong Liu and Hao Chen},
    title = {{Matryoshka}: fuzzing deeply nested branches},
    booktitle = {ACM Conference on Computer and Communications Security (CCS)},
    address = {London, UK},
    year = {2019},
    month = {11},
}

@InProceedings{greyone,
    author = {Shuitao Gan and Chao Zhang and Peng Chen and Bodong Zhao and Xiaojun Qin and Dong Wu and Zuoning Chen},
    title = {{GREYONE}: Data Flow Sensitive Fuzzing},
    booktitle = {{USENIX} Security Symposium ({USENIX} Security)},
    year={2020},
    address = {Boston, MA},
    month = aug,
}

@inproceedings{qsym,
  title={{QSYM}: A practical concolic execution engine tailored for hybrid fuzzing},
  author={Yun, Insu and Lee, Sangho and Xu, Meng and Jang, Yeongjin and Kim, Taesoo},
  booktitle={27th {USENIX} Security Symposium ({USENIX} Security 18)},
  pages={745--761},
  year={2018}
}

@inproceedings{driller,
  author = {Stephens, Nick and Grosen, John and Salls, Christopher and Dutcher, Andrew and Wang, Ruoyu and Corbetta, Jacopo and Shoshitaishvili, Yan and Kruegel, Christopher and Vigna, Giovanni},
  title = {Driller: augmenting fuzzing through selective symbolic execution},
  booktitle = {Proceedings of the Network and Distributed System Security Symposium},
  year = {2016}
}

@inproceedings{babic2019fudge,
  title={Fudge: fuzz driver generation at scale},
  author={Babi{\'c}, Domagoj and Bucur, Stefan and Chen, Yaohui and Ivan{\v{c}}i{\'c}, Franjo and King, Tim and Kusano, Markus and Lemieux, Caroline and Szekeres, L{\'a}szl{\'o} and Wang, Wei},
  booktitle={Proceedings of the 2019 27th ACM Joint Meeting on European Software Engineering Conference and Symposium on the Foundations of Software Engineering},
  pages={975--985},
  year={2019}
}

@inproceedings{ispoglou2020fuzzgen,
  title={{FuzzGen}: Automatic Fuzzer Generation},
  author={Ispoglou, Kyriakos and Austin, Daniel and Mohan, Vishwath and Payer, Mathias},
  booktitle={29th USENIX Security Symposium (USENIX Security 20)},
  pages={2271--2287},
  year={2020}
}

@inproceedings{zhang2021apicraft,
  title={{APICraft}: Fuzz Driver Generation for Closed-source {SDK} Libraries},
  author={Zhang, Cen and Lin, Xingwei and Li, Yuekang and Xue, Yinxing and Xie, Jundong and Chen, Hongxu and Ying, Xinlei and Wang, Jiashui and Liu, Yang},
  booktitle={30th USENIX Security Symposium (USENIX Security 21)},
  pages={2811--2828},
  year={2021}
}

@inproceedings{green2022graphfuzz,
  author={Green, Harrison and Avgerinos, Thanassis},
  booktitle={2022 IEEE/ACM 44th International Conference on Software Engineering (ICSE)}, 
  title={{GraphFuzz}: Library {API} Fuzzing with Lifetime-aware Dataflow Graphs}, 
  year={2022},
  pages={1070-1081},
}

@inproceedings{jung2021winnie,
  title={Winnie: Fuzzing windows applications with harness synthesis and fast cloning},
  author={Jung, Jinho and Tong, Stephen and Hu, Hong and Lim, Jungwon and Jin, Yonghwi and Kim, Taesoo},
  booktitle={Proceedings of the 2021 Network and Distributed System Security Symposium (NDSS 2021)},
  year={2021}
}

@inproceedings{chen2022jigsaw,
  title={{JIGSAW}: Efficient and Scalable Path Constraints Fuzzing},
  author={Chen, Ju and Wang, Jinghan and Song, Chengyu and Yin, Heng},
  booktitle={2022 IEEE Symposium on Security and Privacy (SP)},
  pages={1531--1531},
  year={2022},
  organization={IEEE Computer Society}
}

@online{cjson,
  title = {cJSON},
  url = {https://github.com/DaveGamble/cJSON}
}

@conference{serebryany2017oss,
  title={{OSS-Fuzz}- {Google}'s continuous fuzzing service for open source software},
  author={Serebryany, Kostya},
  publisher = {USENIX Association},
  year={2017}
}

@online{bindgen,
  title = {rust-bindgen},
  url = {https://github.com/rust-lang/rust-bindgen}
}

@online{bindgen_cpp,
  title = {Generating Bindings to C++},
  url = {https://rust-lang.github.io/rust-bindgen/cpp.html}
}

@inproceedings{e9patch,
  title={Binary rewriting without control flow recovery},
  author={Duck, Gregory J and Gao, Xiang and Roychoudhury, Abhik},
  booktitle={Proceedings of the 41st ACM SIGPLAN Conference on Programming Language Design and Implementation},
  pages={151--163},
  year={2020}
}

@inproceedings{e9afl,
  title={Scalable Fuzzing of Program Binaries with {E9AFL}},
  author={Gao, Xiang and Duck, Gregory J and Roychoudhury, Abhik},
  booktitle={2021 36th IEEE/ACM International Conference on Automated Software Engineering (ASE)},
  pages={1247--1251},
  year={2021},
  organization={IEEE}
}

@online{lpm,
  title = {Structure-Aware Fuzzing with {libFuzzer}},
  url = {https://github.com/google/fuzzing/blob/master/docs/structure-aware-fuzzing.md}
}

@online{fdp,
  title = {How To Split A Fuzzer-Generated Input Into Several},
  url = {https://github.com/google/fuzzing/blob/master/docs/split-inputs.md}
}

@inproceedings{srivastava2021gramatron,
  title={Gramatron: Effective grammar-aware fuzzing},
  author={Srivastava, Prashast and Payer, Mathias},
  booktitle={Proceedings of the 30th ACM SIGSOFT International Symposium on Software Testing and Analysis},
  pages={244--256},
  year={2021}
}

@article{gross2018fuzzil,
  title={Fuzzil: Coverage guided fuzzing for {Javascript} engines},
  author={Gro{\ss}, Samuel},
  journal={Department of Informatics, Karlsruhe Institute of Technology},
  year={2018}
}

@inproceedings{veggalam2016ifuzzer,
  title={Ifuzzer: An evolutionary interpreter fuzzer using genetic programming},
  author={Veggalam, Spandan and Rawat, Sanjay and Haller, Istvan and Bos, Herbert},
  booktitle={European Symposium on Research in Computer Security},
  pages={581--601},
  year={2016},
  organization={Springer}
}

@inproceedings{wang2019superion,
  title={Superion: Grammar-aware greybox fuzzing},
  author={Wang, Junjie and Chen, Bihuan and Wei, Lei and Liu, Yang},
  booktitle={2019 IEEE/ACM 41st International Conference on Software Engineering (ICSE)},
  pages={724--735},
  year={2019},
  organization={IEEE}
}

@inproceedings{lee2020montage,
  title={Montage: A Neural Network Language Model-Guided {JavaScript} Engine Fuzzer},
  author={Lee, Suyoung and Han, HyungSeok and Cha, Sang Kil and Son, Sooel},
  booktitle={29th {USENIX} Security Symposium ({USENIX} Security 20)},
  pages={2613--2630},
  year={2020}
}

@inproceedings{aschermann2019nautilus,
  title={{NAUTILUS}: Fishing for Deep Bugs with Grammars.},
  author={Aschermann, Cornelius and Frassetto, Tommaso and Holz, Thorsten and Jauernig, Patrick and Sadeghi, Ahmad-Reza and Teuchert, Daniel},
  booktitle={Proceedings of the 2019 Network and Distributed System Security Symposium (NDSS)},
  year={2019}
}

@inproceedings{blazytko2019grimoire,
  title={{GRIMOIRE}: Synthesizing structure while fuzzing},
  author={Blazytko, Tim and Bishop, Matt and Aschermann, Cornelius and Cappos, Justin and Schl{\"o}gel, Moritz and Korshun, Nadia and Abbasi, Ali and Schweighauser, Marco and Schinzel, Sebastian and Schumilo, Sergej and others},
  booktitle={28th USENIX Security Symposium (USENIX Security 19)},
  pages={1985--2002},
  year={2019}
}

@inproceedings{chen2021one,
  title={One engine to fuzz’em all: Generic language processor testing with semantic validation},
  author={Chen, Yongheng and Zhong, Rui and Hu, Hong and Zhang, Hangfan and Yang, Yupeng and Wu, Dinghao and Lee, Wenke},
  booktitle={2021 IEEE Symposium on Security and Privacy (SP)},
  pages={642--658},
  year={2021},
  organization={IEEE}
}

@inproceedings{sun2021healer,
  title={{HEALER}: Relation Learning Guided Kernel Fuzzing},
  author={Sun, Hao and Shen, Yuheng and Wang, Cong and Liu, Jianzhong and Jiang, Yu and Chen, Ting and Cui, Aiguo},
  booktitle={Proceedings of the ACM SIGOPS 28th Symposium on Operating Systems Principles},
  pages={344--358},
  year={2021}
}

@inproceedings{pailoor2018moonshine,
  title={{MoonShine}: Optimizing {OS} Fuzzer Seed Selection with Trace Distillation},
  author={Pailoor, Shankara and Aday, Andrew and Jana, Suman},
  booktitle={27th USENIX Security Symposium (USENIX Security 18)},
  pages={729--743},
  year={2018}
}

@inproceedings{atlidakis2019restler,
  title={Restler: Stateful {REST} {API} fuzzing},
  author={Atlidakis, Vaggelis and Godefroid, Patrice and Polishchuk, Marina},
  booktitle={2019 IEEE/ACM 41st International Conference on Software Engineering (ICSE)},
  pages={748--758},
  year={2019},
  organization={IEEE}
}

@article{atlidakis2020pythia,
  title={Pythia: grammar-based fuzzing of {REST APIs} with coverage-guided feedback and learning-based mutations},
  author={Atlidakis, Vaggelis and Geambasu, Roxana and Godefroid, Patrice and Polishchuk, Marina and Ray, Baishakhi},
  journal={arXiv preprint arXiv:2005.11498},
  year={2020}
}

@inproceedings{holler2012fuzzing,
  title={Fuzzing with code fragments},
  author={Holler, Christian and Herzig, Kim and Zeller, Andreas},
  booktitle={21st USENIX Security Symposium (USENIX Security 12)},
  pages={445--458},
  year={2012}
}

@inproceedings{zhong2020squirrel,
  title={Squirrel: Testing database management systems with language validity and coverage feedback},
  author={Zhong, Rui and Chen, Yongheng and Hu, Hong and Zhang, Hangfan and Lee, Wenke and Wu, Dinghao},
  booktitle={Proceedings of the 2020 ACM SIGSAC Conference on Computer and Communications Security},
  pages={955--970},
  year={2020}
}

@inproceedings{wang2017skyfire,
  title={Skyfire: Data-driven seed generation for fuzzing},
  author={Wang, Junjie and Chen, Bihuan and Wei, Lei and Liu, Yang},
  booktitle={2017 IEEE Symposium on Security and Privacy (SP)},
  pages={579--594},
  year={2017},
  organization={IEEE}
}

@inproceedings{you2019profuzzer,
  title={Profuzzer: On-the-fly input type probing for better zero-day vulnerability discovery},
  author={You, Wei and Wang, Xueqiang and Ma, Shiqing and Huang, Jianjun and Zhang, Xiangyu and Wang, XiaoFeng and Liang, Bin},
  booktitle={2019 IEEE symposium on security and privacy (SP)},
  pages={769--786},
  year={2019},
  organization={IEEE}
}

@inproceedings{godefroid2017learn,
  title={Learn\&fuzz: Machine learning for input fuzzing},
  author={Godefroid, Patrice and Peleg, Hila and Singh, Rishabh},
  booktitle={2017 32nd IEEE/ACM International Conference on Automated Software Engineering (ASE)},
  pages={50--59},
  year={2017},
  organization={IEEE}
}

@online{sqlsmith,
  title = {{SQLsmith}},
  url = {https://github.com/anse1/sqlsmith}
}

@online{fuzzgen_libvpx,
    title = {A fuzz driver generated by {FuzzGen} for libvpx},
    url = {https://github.com/HexHive/FuzzGen/tree/master/examples/libvpx}
}

@online{sqlite3_gfuzz_schema,
    title = {A schema for generating sqlite3 fuzz drivers written by {GraphFuzz}},
    url = {https://github.com/hgarrereyn/GraphFuzz/blob/master/experiments/sqlite3/in/f1/schema.yaml}
}

@online{pulling_jepgs,
    title = {Pulling {JPEGs} out of thin air},
    url = {https://lcamtuf.blogspot.com/2014/11/pulling-jpegs-out-of-thin-air.html}
}

@inproceedings{jeong2022utopia,
  title={{UTOPIA}: Automatic Generation of Fuzz Driver using Unit Tests},
  author={Jeong, Bokdeuk and Jang, Joonun and Yi, Hayoon and Moon, Jiin and Kim, Junsik and Jeon, Intae and Kim, Taesoo and Shim, WooChul and Hwang, Yong Ho},
  booktitle={2023 IEEE Symposium on Security and Privacy (SP)},
  pages={746--762},
  year={2022},
  organization={IEEE Computer Society}
}

@InProceedings{miner2023,
  author = {Lyu, Chenyang and Xu, Jiacheng and Ji, Shouling and Zhang, Xuhong and Wang, Qinying and Zhao, Binbin and Pan, Gaoning and Cao, Wei and Chen, Peng and Beyah, Raheem},
  title = {{MINER}: A Hybrid Data-Driven Approach for {REST API} Fuzzing},
  booktitle = {{USENIX} Security Symposium ({USENIX} Security)},
  year={2023},
  address = {Ananeim, CA},
  month = aug,
}

\appendix

\section{Appendix}
\balance

\subsection{DSL Grammar}
\label{sec:dsl_grammar}

\begin{figure}[h]
\begin{align*}
Program & :=  Line \ | \ Line; Program \\ 
Line & := Index \  Statement \\
Statement & := Call \ | \ Load \ | \ File \ | \ Assert \ | \ Update \\
Call & := \mathbf{call} \ Name (ArgList) \\
Load & := \mathbf{load} \ Type = Value \\
Update & := \mathbf{update} \ Index[Fields] = Index \\
Assert & := \mathbf{assert} \ Rule \\
File & := \textbf{file write} \ | \ \textbf{file read} \ Index \\
ArgList & := Index \ | \ Index, ArgList \\
Rule & := \textbf{non\_null}(Index) \  | \ \textbf{eq}(Index, Index) \\
Index & := \mathbf{<}\text{numeric literal}\mathbf{>} \\
Name & := \text{function name} \\
Type & := \text{type name of value} \\
Value & := \text{serialized value} \\
Fields &:= \text{path to locate a value inside struct or array }
\end{align*}
\caption{Grammar of \tool's DSL. }
\label{fig:dsl}
\end{figure}

As shown in \autoref{fig:dsl}, an individual DSL Program consists of a sequence of lines. The index at the beginning of each line is a unique ID (e.g., an incremental number), which can be referenced by other statements. 
Since the values in arguments can have various types, our DSL encodes them accordingly.
Primitive types are easy to convert to and from strings. In the case of composite data types, such as an array or custom struct, we serialize their internal elements, following a structured order similar to the JSON format. 
Specifically, when dealing with long lists of primitive values, we optimize them for efficiency by encoding them in Base64 (as seen in line 0 of \autoref{fig:dsl_example}). However, serializing a pointer is tricky since it may point to a value shared by multiple objects. Therefore, we only serialize the destination to which it points in the DSL program, without including any additional information.

\subsection{Examples of inaccurate fuzz drivers}
\label{sec:wrong_drivers}
\begin{figure}[H]
\begin{lstlisting}[language=c, firstnumber=415, xleftmargin=0.3em]
struct vpx_codec_ctx *ctx_hEP_1;// = &ctx_hEP_0; (*@\label{lst:misuse:val}@*)

// Dependence family #3 Definition
dep_3 = (struct vpx_codec_ctx *)ctx_hEP_1;
// initializing argument 'cfg_Ywn'
struct vpx_codec_dec_cfg cfg_Ywn_0;

*(uint32_t*)((uint64_t)&cfg_Ywn_0 + 0) = (E.eat1() & 0x3f) + 1; /* UNKNOWN */
*(uint32_t*)((uint64_t)&cfg_Ywn_0 + 4) = 0; /* UNKNOWN */
*(uint32_t*)((uint64_t)&cfg_Ywn_0 + 8) = 0; /* UNKNOWN */
struct vpx_codec_dec_cfg *cfg_Ywn_1 = &cfg_Ywn_0;

if (vpx_codec_dec_init_ver(dep_3, dep_6, cfg_Ywn_1, 0, 12)) { /* vertex #4 */  (*@\label{lst:misuse:call}@*)
    return 0;
}
\end{lstlisting}
\caption{An example of misuse of consumer code in \file{libvpx}'s fuzz driver~\cite{fuzzgen_libvpx} generated by FuzzGen.}
\label{fig:misuse_cosumer}
\end{figure}

\autoref{fig:misuse_cosumer} shows a fuzz driver for \file{libvpx} generated automatically by FuzzGen. On Line~\ref{lst:misuse:val}, the value of \lstinline|ctx_hEP_1| is an uninitialized pointer, which is then used directly as an argument in the call on Line~\ref{lst:misuse:call}. In addition, upon examining the fuzz driver for \file{libaom} that was released by the authors of FuzzGen, we found an incompatibility with the newest version of \file{libaom}. Specifically, the initialization of \verb|aom_codec_dec_cfg| objects in the fuzz driver overwrites 17-20 bytes despite the size of the object being reduced to only 16 bytes.

The schema~\cite{sqlite3_gfuzz_schema} for \file{sqlite3} is written by the authors of GraphFuzz. When the program fails to invoke \lstinline|sqlite3_prepare_v2| on line 72, it calls \lstinline|close_all| to release all \lstinline|sqlite3*| pointers. However, other types of allocated resources, such as \lstinline|sqlite3_str*| pointers, may still be leaked. Moreover, the schema does not verify whether \lstinline|sqlite3*| is initialized successfully or not in \lstinline|new_database|. 

\clearpage
\end{document}